\documentclass{arxivStyle}
\usepackage{cite}
\usepackage{amsmath,amssymb,amsfonts}
\usepackage{graphicx,color}
\usepackage{textcomp}
\def\BibTeX{{\rm B\kern-.05em{\sc i\kern-.025em b}\kern-.08em
    T\kern-.1667em\lower.7ex\hbox{E}\kern-.125emX}}
\AtBeginDocument{\definecolor{ojcolor}{cmyk}{0.93,0.59,0.15,0.02}}
\usepackage{acronym}
\usepackage{amsmath}
\usepackage{tabularx,booktabs}
\usepackage{multicol}
\usepackage{makecell}
\usepackage{amsfonts}
\usepackage{bm}
\usepackage{tablefootnote}
\usepackage{url}
\usepackage{caption}
\usepackage{amssymb}
\usepackage{amsthm}
\usepackage{chngcntr}
\usepackage[table]{xcolor}
\usepackage{array}        
\usepackage{array, multirow}
\usepackage{caption}      
\usepackage{graphicx}
\usepackage{diagbox}
\usepackage{float}
\usepackage{comment} 
\usepackage{placeins}
\usepackage{tikz}
\usepackage{graphicx} 
\usepackage{adjustbox}

\usepackage{tikz}
\usepackage{pgfplots}
\usepgfplotslibrary{groupplots}
\usepgfplotslibrary{fillbetween}
\pgfplotsset{compat=1.18}

\usepackage[caption=false,font=normalsize]{subfig}
\usetikzlibrary{arrows.meta, positioning, shapes.geometric}
\usepackage{array}
\newcolumntype{C}{>{\centering\arraybackslash}X}

\usepackage{optidef}

\definecolor{better}{rgb}{0.7, 0.85, 1}
\definecolor{worse}{rgb}{1, 0.9, 0.6}    
\definecolor{neutral}{rgb}{1, 1, 1}
\definecolor{diaggray}{gray}{0.85}
\newcolumntype{Y}{>{\centering\arraybackslash}X}
\newcolumntype{?}{!{\vrule width 1.5pt}}
\usepackage[numbers,sort&compress]{natbib}
\acrodef{CAV}[CAV]{connected and autonomous vehicle}
\acrodef{V2I}[V2I]{vehicle-to-infrastructure}
\acrodef{V2V}[V2V]{vehicle-to-vehicle}
\acrodef{V2N}[V2N]{vehicle-to-network}
\acrodef{V2X}[V2X]{vehicle-to-everything}
\acrodef{MAP}[MAP]{multi-agent planning}
\acrodef{MINLP}[MINLP]{mixed-integer non-linear program}
\acrodef{FIFO}[FIFO]{first-in-first-out}
\acrodef{PMF}[PMF]{probability mass function}
\acrodef{SUMO}[SUMO]{Simulation of Urban MObility}
\acrodef{IDM}[IDM]{Intelligent Driver Model}
\acrodef{EIDM}[EIDM]{Extended IDM}
\acrodef{CACC}[CACC]{Cooperative Adaptive Cruise Control}
\acrodef{TSC}[TSC]{traffic signal control}
\acrodef{TUC}[TUC]{traffic-responsive urban control}
\acrodef{LQ}[LQ]{linear-quadratic}
\acrodef{MPC}[MPC]{model predictive control}
\acrodef{LTM}[LTM]{link transmission model}
\acrodef{RL}[RL]{reinforcement learning}
\acrodef{MARL}[MARL]{multi-agent reinforcement learning}
\acrodef{SAC}[SAC]{soft actor-critic}
\acrodef{SURTRAC}[SURTRAC]{Scalable Urban Traffic Control}
\acrodef{MP}[MP]{max-pressure}
\acrodef{AIM}[AIM]{autonomous intersection management}
\acrodef{ILP}[ILP]{integer linear programming}
\acrodef{TLS}[TLS]{traffic light signals}
\acrodef{ETSI}[ETSI]{European telecommunications standards institute}
\acrodef{SAE}[SAE]{society of automotive engineering}

\theoremstyle{definition}

\usepackage{algorithm}
\usepackage{algpseudocode}
\algnewcommand{\LeftComment}[1]{\Statex \(\triangleright\) \textit{#1}}

\usepackage{soul,color}

\usepackage{ragged2e}
\usepackage{blindtext}

\def\authorrefmark#1{\ensuremath{^{\textbf{#1}}}}

\begin{document}

\title{Decentralized Multi-Agent Urban Traffic Management via Spatio-Temporal Mobility Profile Planning} %

\author{~Lorenzo~Mario~Amorosa\authorrefmark{1}~(Member,~IEEE), Lorenzo~Farina\authorrefmark{1} (Student~Member,~IEEE), Vittorio~Todisco\authorrefmark{1} (Member,~IEEE), and~Alessandro~Bazzi\authorrefmark{1}~(Senior~Member,~IEEE)}
\affil{Department of Electrical, Electronic and Information Engineering (DEI), ``Guglielmo Marconi", University of Bologna \& \\WiLab - National Wireless Communications Laboratory of CNIT, Bologna, Italy.}
\corresp{Corresponding author: Lorenzo~Mario~Amorosa (email: lorenzomario.amorosa@unibo.it).}

\begin{abstract}
As modern cities face increasingly severe traffic congestion, connected and autonomous vehicles (CAVs) have emerged as a crucial enabling technology for next-generation intelligent traffic management. However, fully realizing this potential is hindered by the limitations of current paradigms. Existing approaches typically optimize localized interactions rather than system-wide efficiency, incur severe communication overhead, or lack the deterministic guarantees required for safe kinematic execution. Furthermore, current multi-agent adaptations are frequently restricted to small predefined scenarios, failing to scale across large and complex urban networks. To bridge this gap, this paper introduces VeloCity, a decentralized multi-agent spatio-temporal mobility profile planning framework designed for CAVs operating in arbitrary urban areas. To minimize vehicles' travel times, VeloCity distributes mobility profile optimization directly to individual CAVs. Vehicles query a localized traffic coordinator for a reservation table, independently compute their fastest conflict-free mobility profile, and reserve their requested space-time slots back with the coordinator. By natively adapting to any arbitrary road topology, the framework manages highly irregular urban areas without requiring scenario-specific tuning, all while guaranteeing collision-free and physically executable vehicle trajectories. Extensive simulations across four large-scale real-world urban maps (Tokyo, Manhattan, Rome, and Bologna) demonstrate the framework's scalability. Compared to established state-of-the-art models, VeloCity yields drastically lower travel times, tightly bounds delay variance, and successfully prevents congestion gridlocks even under extremely high vehicular densities.
\end{abstract}



\maketitle

\section{INTRODUCTION}
\label{sec:introduction}

\IEEEPARstart{A}{s} urban traffic complexity grows, traditional traffic management paradigms are facing unprecedented scalability challenges \cite{10004534}. To overcome these limitations, \acp{CAV} have emerged as a crucial enabling technology for next-generation intelligent transportation. 
Conventional traffic control systems, such as fixed or adaptive traffic signals, often fail to exploit the fine-grained coordination and communication capabilities enabled by emerging \ac{V2I} networks \cite{farina2026v2nbasedalgorithmcommunicationprotocol}.
To mitigate these inefficiencies, advanced car-following models driven by both local and collective perception have been proposed to significantly improve traditional traffic flow management \cite{10547481}. However, because they predominantly optimize for localized interactions rather than system-wide efficiency, their scalability remains constrained. Consequently, as urban centers grow more complex, there is a clear need for flexible and decentralized traffic planning frameworks capable of managing high-density vehicle coordination across arbitrary road topologies without relying on rigid scenario-specific control infrastructure.
{\Ac{MAP}} offers a promising formulation for this challenge, treating each vehicle as an intelligent agent tasked with finding collision-free paths from a start to a target location \cite{dresner2008multiagent}. However, existing traffic-oriented MAP adaptations often constrain themselves to small and predefined scenarios, such as single intersections or simple merging lanes, failing to generalize to arbitrary map areas featuring diverse topological elements like roundabouts and multi-leg intersections \cite{abbas2023autonomous, 8585057}. 

The practical deployment of advanced \ac{CAV} coordination relies intrinsically on accurate environment perception and the reliable execution of planned mobility profiles over \ac{V2X} networks. Recently, standardization organizations such as the \ac{ETSI} and the \ac{SAE} are actively establishing the necessary foundations to make this feasible. Building upon well-established cooperative awareness frameworks validated through both simulations \cite{avino2018support} and field testing \cite{rapelli2024oscar}, recent standardization efforts have expanded toward enabling complex maneuver coordination. Specifically, \ac{SAE} has formalized the Maneuver Sharing and Coordinating Service in its J3186 and J3216 standards \cite{SAEJ3186,SAEJ3216}. Similarly, \ac{ETSI} is detailing analogous specifications for maneuver coordination \cite{etsi103.578,etsi103.561}. These specifications dictate the precise messages and communication protocols required for vehicles to safely negotiate maneuvers in shared spaces, such as intersections. 
Importantly, these standards specify \emph{how} vehicles exchange information for maneuver coordination, but they do not prescribe specific conflict-resolution or traffic-management algorithms. 
This decoupling creates a critical gap that must be filled by application developers, directly motivating the need for robust and decentralized planners capable of supplying the optimized maneuver strategies that these protocols will carry.

To bridge this gap, this paper introduces \emph{VeloCity}, a decentralized multi-agent spatio-temporal mobility profile planning framework for \acp{CAV} operating in arbitrary urban areas. In our framework, each map area is equipped with a local controller that maintains a spatio-temporal reservation table for its respective road segments. Crucially, instead of the controller solving a joint optimization problem for all agents, mobility profile optimization is distributed across the vehicles. Upon approaching a managed area, a vehicle queries the controller for the current reservation table, locally computes its fastest conflict-free mobility profile, and subsequently reserves its requested spatio-temporal slots back at the controller. This decentralized \ac{MAP} approach aims at accommodating arbitrary road geometry, including complex intersections and roundabouts, without requiring scenario-specific tuning. 
Specifically, the main contributions of this paper are summarized as follows:
\begin{itemize}
\item We propose \emph{VeloCity}, a decentralized \ac{MAP} framework designed for urban traffic management. By decomposing the mobility optimization problem into sequential spatio-temporal planning tasks to be executed at each \ac{CAV}, the framework guarantees collision-free and physically executable vehicle trajectories. Its flexible space-time reservation architecture intrinsically supports arbitrary and heterogeneous road topologies without requiring scenario-specific tuning.
\item We conduct an extensive performance evaluation utilizing the \ac{SUMO} simulator \cite{sumoRef} across four diverse, large-scale urban environments imported directly from real-world OpenStreetMap\footnote{See: \url{https://www.openstreetmap.org/}} data (Tokyo, Manhattan, Rome, and Bologna). Through extensive experiments, we demonstrate that VeloCity significantly outperforms established baselines. The proposed approach yields drastically lower travel times, tightly bounded delay variance, and robustness against congestion gridlocks even under very high vehicular densities.
\item To ensure reproducibility and support the broader research community, we release the complete open-source implementation of VeloCity. The repository, which includes the core algorithms, the four large-scale urban scenarios, and all simulation scripts, is publicly available at {\url{https://github.com/V2Xgithub/VeloCity}}.
\end{itemize}
To the best of our knowledge, VeloCity represents the first \ac{V2X}-based framework to successfully scale decentralized and deterministic mobility profile optimization to arbitrary city-wide road networks.

The remainder of this paper is organized as follows. Section~\ref{sec:related_work} reviews the relevant literature. Section~\ref{sec:system_model} defines the system model and formulates the multi-agent mobility profile optimization problem. Section~\ref{sec:methodology} details the proposed VeloCity methodology. Section~\ref{sec:experiments} presents the experimental setup and discusses the numerical results. Finally, Section~\ref{sec:conclusion} summarizes the concluding remarks and outlines potential directions for future work.

\section{RELATED WORK}
\label{sec:related_work}

The management of traffic flows in urban road networks has long been a critical challenge for transportation engineering. For decades, the primary mechanism for resolving vehicle conflicts at junctions has been the traditional traffic light, with researchers focusing heavily on phase optimization. While early traffic signal control systems utilized offline static plans or heuristic adaptive strategies like SCOOT and SCATS~\cite{gartner1975optimization, stevanovic2009scoot}, modern approaches have embraced predictive control, such as \ac{MPC}~\cite{aboudolas2010rolling, wei2025hierarchical}, schedule-driven methods~\cite{smith2013smart, hu2021incorporating}, and max-pressure control~\cite{varaiya2013max, xu2024smoothing}. More recently, a hybrid architecture exploiting both \ac{MPC} and \ac{RL} \cite{remmerswaal2022combined} and \ac{MARL} frameworks have also been proposed to manage local networks~\cite{ma2023learning, wang2025towards}. 
However, regardless of the underlying mathematical formulation, these solutions remain fundamentally constrained by the traffic-light paradigm and therefore cannot fully exploit the coordination capabilities enabled by advances in vehicle automation and \ac{V2X} connectivity.

Recently, \acp{CAV} encouraged a paradigm shift toward \ac{AIM}, where replacing traffic lights with \ac{V2X}-based maneuver coordination and dynamic space-time reservations can eliminate stop-and-go delays and drastically increase intersection capacity~\cite{tachet2016revisiting}.  Numerous \ac{AIM} protocols have emerged~\cite{maksimovski2021survey,guo2019urban}. Early decentralized architectures functioned as virtual traffic lights, managing crossing orders via sequential \ac{V2V} priority passing~\cite{bazzi2016distributed}, clustering~\cite{jame2022reducing}, directed collision graphs~\cite{lin2019graph}, or \ac{RL}~\cite{karthikeyan2022autonomous}. While successful at conflict sequencing, these methods do not control kinematic profiles to prevent stop-and-go maneuvers. To fully exploit \ac{AIM}, subsequent research combined scheduling with precise trajectory optimization. For instance, centralized \ac{MPC} and independent energy-minimization controllers were developed to calculate stop-free acceleration profiles based on space-time reservations~\cite{dresner2008multiagent, zhang2019decentralized, hult2018energy}. A specific subset of this literature bridges these models with real-world communication constraints, such as formulating protocols for packet loss and 4G/5G latency~\cite{zheng2017delay, lu2021optimization, farina2026v2nbasedalgorithmcommunicationprotocol}. Yet, while these single-intersection frameworks successfully optimize local vehicle trajectories, they do not address how such solutions scale to larger arbitrary road networks.

Translating these approaches to large-scale urban grids introduces severe computational and communication bottlenecks. To date, efforts to extend \ac{AIM} typically link a limited number of adjacent managers rather than addressing city-scale grids. For instance, expanding reservation architectures via time-based A* searches can suffer from information staleness, as vehicles route themselves using instantaneously outdated congestion data~\cite{hausknecht2011autonomous}. To address this, dynamic market-inspired auctions~\cite{vasirani2012market} and consensus protocols~\cite{wuthishuwong2017consensus} have been proposed; however, both introduce severe communication overhead when scaled to large road scenarios. To explicitly consider time-evolution and precise kinematics, \cite{okoso2021network} reformulated routing as a time-expanded network flow problem, though its reliance on \ac{ILP} necessitates a fully centralized and computationally heavy architecture. Recently, alternative solutions have turned to \ac{MARL} by modeling road segments or vehicles as intelligent agents~\cite{tranos2024large, liu2025large}. Beyond reintroducing significant communication overhead, these \ac{MARL} policies remain fundamentally reactive and rely on opaque neural networks, precluding the exact determinism and formal kinematic safety guarantees required for autonomous traffic coordination. 

Consequently, a significant gap remains for a truly scalable \ac{AIM} architecture capable of handling high-density arbitrary urban networks. The literature currently lacks an adaptive framework that resolves coordination through purely decentralized and physically executable optimizations, which is a gap that VeloCity aims to fill.

\begin{figure*}[!t]
\centering

\input{figures/Schemes/figure1_FULL}

\caption{(a) Considered urban area, where the red vehicle $i$ approaches the coordinated area: (1) the coordinator provides the reservation table $\mathcal{R}$ to $i$, (2) $i$ locally computes its feasible mobility profile, and (3) $i$ transmits its reservation request $\mathcal{Q}_i$ to the coordinator. (b) Discrete spatial representation of the reservation table $\mathcal{R}(t)$. The road network is mapped to a spatial grid, where solid colored cells represent the vehicles' physical footprints $\mathcal{F}_i(s_i(t))$ and lighter cells indicate their next intended spatial reservations for a $t' > t$.}
\label{fig:full_figure}
\end{figure*}

\section{SYSTEM MODEL AND PROBLEM FORMULATION}
\label{sec:system_model}





{We }consider an urban area managed by a traffic coordinator, including a set of \acp{CAV}\footnote{We use the terms CAV and agent interchangeably.} denoted by $\mathcal{V} = \{1, 2, \dots, N\}$. Each vehicle $i \in \mathcal{V}$ enters the area at a specific requested time $t_{\mathrm{in}, i}$ with a predetermined\footnote{VeloCity focuses on mobility profile optimization given a fixed path; however, it remains fully compatible with proactive dynamic routing strategies. In our simulations, vehicle routing is managed autonomously by \ac{SUMO}.} trajectory $\mathcal{P}_i$ and initial speed $v_{\mathrm{in}, i}$. 
The objective of each CAV is to optimize its mobility profile, i.e., its speed and acceleration over time, to drive and reach the intended destination minimizing its travel time. 
A CAV's trajectory $\mathcal{P}_i$ is represented by a one-dimensional path of total physical length $D_i$, where the placement of vehicle $i$ at time step $t$ is defined by $s_i(t) \in [0, D_i]$. 
This 1D path maps directly to the {two-dimensional} Euclidean road plane\footnote{For multi-level infrastructures, such as bridges or overpasses, vertically overlapping road segments are mapped to logically distinct two-dimensional planes to prevent false spatial conflicts.} via a network geometry function $\gamma_i: [0, D_i] \to \mathbb{R}^2$, which defines the spatial coordinates of the vehicle's reference center. To account for physical dimensions, vehicle $i$ is characterized by a two-dimensional bounding area $\mathcal{F}_i(s_i(t)) \subset \mathbb{R}^2$. This footprint represents the physical space occupied by the vehicle on the map at time step $t$, constructed by sweeping its length $L_i$ and width $W_i$ around the reference center $\gamma_i(s_i(t))$.  
Importantly, each $\mathcal{P}_i$ includes routing, comprising specific vehicle's lane assignments and intersection crossings; however, it does not specify the vehicle's mobility profile, leaving its speed and acceleration over time to be determined. 


\subsection{VEHICLE'S MOBILITY PROFILE}

Each vehicle $i \in \mathcal{V}$ must determine its mobility profile over time, governed by its speed $v_i(t)$ and acceleration $a_i(t)$. To ensure physically executable and safe trajectories, these control variables are strictly bounded by both vehicle-inherent kinematic capabilities and area-specific speed regulations. First, the acceleration is limited by the vehicle's mechanical capabilities:
\begin{equation}
\label{eq:acc_limit}
    a_{\min, i} \leq a_i(t) \leq a_{\max, i}
\end{equation}
where $a_{\min, i}$ and $a_{\max, i}$ represent the maximum deceleration and acceleration constants for vehicle $i$, respectively. Second, the maximum permissible speed depends not only on the vehicle's top performance but also on the localized speed limits of the specific road segment or lane it currently occupies. Let $V_{\max}(\mathbf{p}) : \mathbb{R}^2 \to \mathbb{R}^+$ be a function mapping a 2D coordinate to the local legal speed limit. The speed profile is thus constrained by:
\begin{equation}
\label{eq:speeed_limit}
    0 \leq v_i(t) \leq \min \left( V_{\max, i}, \, V_{\max}\big(\gamma_i(s_i(t))\big) \right)
\end{equation}
where $V_{\max, i}$ is the maximum structural speed of vehicle $i$. By evaluating $V_{\max}$ at the vehicle's instantaneous 2D coordinate $\gamma_i(s_i(t))$, this formulation dynamically enforces area-specific speed constraints, such as speed reductions at sharp turns, roundabouts, or intersection approaches.

\subsection{TRAFFIC COORDINATOR}

The coordination of \acp{CAV} within the managed urban area is supervised by a localized traffic infrastructure coordinator. This coordinator does not solve a joint centralized problem for determining \acp{CAV} profiles. Instead, it serves as a reference point to maintain a spatio-temporal reservation table, denoted by $\mathcal{R}$, while \acp{CAV} optimize their profiles in a decentralized way. The reservation table $\mathcal{R}(t)$ is discretized in time with a resolution $\Delta t$ and maps the {two-dimensional} road network into a spatial grid with a resolution $\Delta s$. At any discrete time step $t$, $\mathcal{R}(t) \subset \mathbb{R}^2$ represents the set of all spatial coordinates reserved by authorized {agents}. When a vehicle $i$ approaches the managed area, it queries the coordinator to retrieve the current state of $\mathcal{R}$. Given $\mathcal{R}(t)$, vehicle $i$ locally computes a feasible mobility profile by ensuring its intended footprint sequence $\mathcal{Q}_i = \{\mathcal{F}_i(s_i(t)),\,\,\, \forall t \geq t_{\mathrm{in}, i}\}$ does not conflict with existing reservations:
\begin{equation}
\label{eq:constr_reserv}
    \mathcal{F}_i(s_i(t)) \cap \mathcal{R}(t) = \emptyset, \quad \forall t \geq t_{\mathrm{in}, i}\,.
\end{equation}
Upon computing this optimal and feasible profile, the vehicle submits its footprint sequence to the coordinator. The coordinator then commits the sequence and updates the reservation table by appending the vehicle's footprint: $\mathcal{R}(t) \leftarrow \mathcal{R}(t) \cup \mathcal{Q}_i$, where $\mathcal{Q}_i$ corresponds to a reservation request, comprising the set of spatial cells occupied by vehicle $i$ over time. {A depiction of the coordinated reservation is presented in Fig.~\ref{fig:full_figure}.



\subsection{PROBLEM FORMULATION}
\label{sec:problem_formulation}

The primary objective of the traffic management framework is to optimize the overall traffic efficiency across the managed urban area while guaranteeing collision-free navigation. We frame this task as a multi-agent optimization problem aimed at minimizing the total travel time of all vehicles traversing the network. To formalize this objective, let $t_{\mathrm{out}, i}$ denote the arrival time step at which vehicle $i \in \mathcal{V}$ successfully reaches the end of its intended path $\mathcal P_i$, satisfying $s_i(t_{\mathrm{out}, i}) \geq D_i$. Given trajectory $\mathcal{P}_i$ and initial speed $v_{\mathrm{in}, i}$ for all vehicles $i \in \mathcal{V}$, the optimization problem can be formulated as a \ac{MINLP}, where the decision variables are $\mathbf{a} = \{a_i(t)\}_{\forall i \in \mathcal{V}}$:

\begin{equation}
\min_{\mathbf{a}} \sum_{i \in \mathcal{V}} \left( t_{\mathrm{out}, i} - t_{\mathrm{in}, i} \right)
\label{eq:objective}
\end{equation}

\noindent subject to the \eqref{eq:acc_limit}, \eqref{eq:speeed_limit}, and \eqref{eq:constr_reserv} constraints for all $i \in \mathcal{V}$ and $t \geq t_{\mathrm{in}, i}$.
In this formulation, \eqref{eq:objective} minimizes the aggregate system-wide delay, representing the total time elapsed between each vehicle's requested entry time $t_{\mathrm{in}, i}$ and its final destination arrival $t_{\mathrm{out}, i}$. 
The optimization problem defined by \eqref{eq:objective} is non-convex and highly complex, hence computing a global joint solution simultaneously would not be practical. This computational challenge motivates the need for a decentralized solution, which we propose in Section~\ref{sec:methodology}.

\section{THE VELOCITY FRAMEWORK}
\label{sec:methodology}

To resolve the global coordination problem formulated in Section~\ref{sec:system_model}.\ref{sec:problem_formulation}, we propose VeloCity: a framework that decomposes the multi-agent optimization problem in \eqref{eq:objective} into a sequence of distributed spatio-temporal planning tasks. 
%
%
%
The core principle in VeloCity is to let each vehicle independently compute its own
feasible mobility profile while coordinating only through a shared
spatio-temporal reservation table $\mathcal{R}$ maintained by the traffic coordinator. The remainder of the section detail the algorithms, which are presented in Algorithm~\ref{alg:velocity_main} and Algorithm~\ref{alg:astar}. 

\begin{algorithm}[t]
\caption{Coordinator Loop}
\label{alg:velocity_main}
\footnotesize
\begin{algorithmic}[1]
\Require Reservation table $\mathcal{R} \gets \emptyset$
\For{each vehicle $i$ entering the coordination area at $t_{\mathrm{in},i}$, with speed $v_{\mathrm{in}, i}$, trajectory $\mathcal{P}_i$, and dynamics bound  $a_{\min,i},a_{\max,i},V_{\max,i}$}
    \State Retrieve current reservation table $\mathcal{R}$
    \State $\mathcal{Q}_i \gets \textsc{LocalPlan}(\mathcal{R}, t_{\mathrm{in},i}, v_{\mathrm{in}, i}, \mathcal{P}_i, a_{\min,i},a_{\max,i},V_{\max,i})$ 
    \State $\mathcal{R} \gets \mathcal{R} \cup \mathcal{Q}_i$ \Comment{Commit trajectory}
\EndFor
\end{algorithmic}
\end{algorithm}

\begin{algorithm}[t]
\caption{\textsc{LocalPlan}}
\label{alg:astar}
\footnotesize
\begin{algorithmic}[1]
\Require Reservation table $\mathcal{R}$, entry time $t_{\mathrm{in},i}$, speed $v_{\mathrm{in}, i}$, trajectory $\mathcal{P}_i$, and dynamics bounds $a_{\min,i},a_{\max,i},V_{\max,i}$
\Ensure Reservation request $\mathcal{Q}_i$
\State $\mathbf{x}_i^0 \gets (s_i^0, v_i^0, t_{\mathrm{in},i})$
\State $g(\mathbf{x}_i^0) \gets 0$
\State $f(\mathbf{x}_i^0) \gets h(\mathbf{x}_i^0)$ \Comment{Eq.~\eqref{eq:astar_heuristic}}
\State $\mathrm{Open} \gets \{\mathbf{x}_i^0\}$ \Comment{priority queue ordered by $f(\cdot)$}
\State $\mathrm{Closed} \gets \emptyset$
\While{$\mathrm{Open} \neq \emptyset$}
    \State $\mathbf{x}_i(t) \gets \arg\min_{\mathbf{x} \in \mathrm{Open}} f(\mathbf{x})$
    \State remove $\mathbf{x}_i(t)$ from $\mathrm{Open}$, add to $\mathrm{Closed}$
    \If{$s_i(t) \geq D_i$} \Comment{goal test}
        \State $\Pi_i \gets$ backtrack path to $\mathbf{x}_i^0$
        \State Convert $\Pi_i$ into reservation request $\mathcal{Q}_i = \{(\mathcal{F}_i^k, t_k)\}_{k=0}^{M_i}$
        \State \Return $\mathcal{Q}_i$
    \EndIf
    \For{each admissible $a_i(t) \in [a_{\min,i}, a_{\max,i}]$}
        \State $v_i(t+\Delta t) \gets v_i(t) + a_i(t)\Delta t$ \Comment{Eq.~\eqref{eq:velocity_update}}
        \If{$0 \le v_i(t+\Delta t) \le \min(V_{\max,i}, V_{\max}(\gamma_i(s_i(t))))$}
            \State $s_i(t+\Delta t) \gets s_i(t) + v_i(t+\Delta t)\Delta t$ \Comment{Eq.~\eqref{eq:position_update}}
            \State $\mathbf{x}_i(t+\Delta t) \gets (s_i(t+\Delta t), v_i(t+\Delta t), t+\Delta t)$
            \If{$\mathcal{F}_i(\mathbf{x}_i(t+\Delta t)) \cap \mathcal{R}(t+\Delta t) = \emptyset$} \Comment{Eq.~\eqref{eq:reservation_check}}
                \If{$\mathbf{x}_i(t+\Delta t) \notin \mathrm{Closed}$}
                    \State $g' \gets g(\mathbf{x}_i(t)) + \Delta t$
                    \If{$\mathbf{x}_i(t+\Delta t) \notin \mathrm{Open}$ \textbf{or} $g' < g(\mathbf{x}_i(t+\Delta t))$}
                        \State $g(\mathbf{x}_i(t+\Delta t)) \gets g'$
                        \State $f(\mathbf{x}_i(t+\Delta t)) \gets g' + h(\mathbf{x}_i(t+\Delta t))$
                        \State parent$(\mathbf{x}_i(t+\Delta t)) \gets \mathbf{x}_i(t)$ \Comment{to compute $\Pi_i$}
                        \State insert/update $\mathbf{x}_i(t+\Delta t)$ in $\mathrm{Open}$
                    \EndIf
                \EndIf
            \EndIf
        \EndIf
    \EndFor
\EndWhile
\State \Return \textsc{infeasible}
\end{algorithmic}
\end{algorithm}
\begin{figure}[!t]
    \centering
    \noindent\resizebox{0.98\columnwidth}{!}{%
        \definecolor{managed}{HTML}{DFF0DC}%
\definecolor{roadcorridor}{HTML}{E9EDEF}%
\definecolor{roadfill}{HTML}{CFD5D9}%
\definecolor{roadedge}{HTML}{8F989F}%
\definecolor{blockfill}{HTML}{EEE8DF}%
\definecolor{blockstroke}{HTML}{C9C0B5}%
\definecolor{courtfill}{HTML}{F8F5EF}%
\definecolor{courtstroke}{HTML}{D9D1C7}%
\definecolor{gridgray}{HTML}{3F4D55}%
\definecolor{pathblue}{HTML}{2563EB}%
\definecolor{currentred}{HTML}{D32F2F}%
\definecolor{futureorange}{HTML}{EF6C00}%
\definecolor{textdark}{HTML}{263238}%
\definecolor{subtext}{HTML}{4B5563}%
\definecolor{boxstroke}{HTML}{DBE5EF}%
\definecolor{legendborder}{HTML}{C9D0D4}%
\definecolor{legendsep}{HTML}{E4E8EB}%
\tikzset{
  roadedge/.style={draw=roadedge,line width=.45pt},
  lane/.style={draw=white,line width=.95pt,dash pattern=on 4.5pt off 3.5pt},
  sample/.style={circle,draw=pathblue,fill=white,line width=.95pt,inner sep=1.8pt},
  labelbox/.style={fill=white,draw=boxstroke,line width=.4pt,rounded corners=1.2pt,inner xsep=3pt,inner ysep=1.6pt},
  legendtext/.style={font=\sffamily\fontsize{8.5}{10}\selectfont,text=textdark},
  legendsub/.style={font=\sffamily\fontsize{7.5}{9}\selectfont,text=subtext},
}
\begin{tikzpicture}[x=0.018cm,y=-0.018cm,font=\sffamily]
  \fill[managed] (37.83,30.98) rectangle (659.30,468.98);

  \foreach \x in {67.3,232.3,397.3,562.3}{\fill[roadcorridor] (\x,50.98) rectangle ++(68,398);}
  \foreach \y in {50.98,215.98,380.98}{\fill[roadcorridor] (67.3,\y) rectangle ++(563,68);}

  \foreach \x in {74.3,239.3,404.3,569.3}{\fill[roadfill] (\x,57.98) rectangle ++(54,384);}
  \foreach \y in {57.98,222.98,387.98}{\fill[roadfill] (74.3,\y) rectangle ++(549,54);}

  \foreach \x in {74.3,128.3,239.3,293.3,404.3,458.3,569.3,623.3}{\draw[roadedge] (\x,57.98)--(\x,441.98);}
  \foreach \x in {101.3,266.3,431.3,596.3}{\draw[lane] (\x,84.98)--(\x,430);}
  \foreach \y in {57.98,112.03,222.98,276.98,387.98,441.98}{\draw[roadedge] (74.3,\y)--(623.3,\y);}
  \foreach \y in {84.98,249.98,414.98}{\draw[lane] (100.45,\y)--(603.3,\y);}

  \foreach \x in {146.3,311.3,476.3}{
    \foreach \y in {129.98,294.98}{
      \draw[fill=blockfill,draw=blockstroke,line width=.55pt,rounded corners=3.5pt] (\x,\y) rectangle ++(75,75);
    }
  }
  \draw[fill=courtfill,draw=courtstroke,line width=.4pt,rounded corners=2pt] (172.55,156.23) rectangle ++(22.5,22.5);
  \draw[fill=courtfill,draw=courtstroke,line width=.4pt,rounded corners=2pt] (337.55,321.23) rectangle ++(22.5,22.5);
  \draw[fill=courtfill,draw=courtstroke,line width=.4pt,rounded corners=2pt] (502.55,156.23) rectangle ++(22.5,22.5);

  \begin{scope}
    \clip (74.3,57.98) rectangle ++(54,384)
          (239.3,57.98) rectangle ++(54,384)
          (404.3,57.98) rectangle ++(54,384)
          (569.3,57.98) rectangle ++(54,384)
          (74.3,57.98) rectangle ++(549,54)
          (74.3,222.98) rectangle ++(549,54)
          (74.3,387.98) rectangle ++(549,54);
    \foreach \x in {84,102,...,624}{
      \foreach \y in {66,84,...,444}{
        \fill[gridgray,opacity=.56] (\x,\y) circle[radius=1.2];
      }
    }
  \end{scope}

  \draw[white,line width=7pt,line cap=round,line join=round]
    (122.4,414.98)--(266.3,414.98)--(266.3,249.98)--(431.3,249.98)--(431.3,84.98)--(566.3,84.98);
  \draw[pathblue,line width=3.5pt,line cap=round,line join=round,-{Triangle[length=8pt,width=10pt]}]
    (122.4,414.98)--(266.3,414.98)--(266.3,249.98)--(431.3,249.98)--(431.3,84.98)--(566.8,84.98);

  \foreach \p in {
  (122.4,414.98),(186.9,414.98),(251.5,414.98),
                    (266.3,365.28),(266.3,300.78),(280.1,249.98),(344.6,249.98),
                    (409.1,249.98),(431.3,207.58),(431.3,143.08),(437.7,84.98),
                    (502.2,84.98),(566.8,84.98)
                    }{
    \node[sample] at \p {};
  }

  \draw[fill=currentred,fill opacity=.12,draw=currentred,line width=1.1pt,dash pattern=on 3.5pt off 2.5pt,rounded corners=5.5pt]
    (241.3,335.98) rectangle ++(50,78);
  \draw[fill=futureorange,fill opacity=.12,draw=futureorange,line width=1.1pt,dash pattern=on 3.5pt off 2.5pt,rounded corners=5.5pt]
    (406.3,170.98) rectangle ++(50,78);

  \draw[fill=currentred,draw=textdark,line width=.7pt,rounded corners=2.7pt]
    (256.3,356.98) rectangle ++(20,36);
  \draw[fill=cyan!12,draw=gray!70,line width=.45pt,rounded corners=1.4pt]
    (260.3,366.98) rectangle ++(12,16);
  \draw[fill=futureorange,draw=textdark,line width=.7pt,rounded corners=2.7pt]
    (421.3,191.98) rectangle ++(20,36);
  \draw[fill=cyan!12,draw=gray!70,line width=.45pt,rounded corners=1.4pt]
    (425.3,201.98) rectangle ++(12,16);

  \node[labelbox,text=pathblue,font=\bfseries\fontsize{8.5}{10}\selectfont] at (100,415) {$s_i^0$};
  \node[labelbox,text=pathblue,font=\bfseries\fontsize{8.5}{10}\selectfont] at (220,369) {$s_i^{k'}$};
  \node[labelbox,text=pathblue,font=\bfseries\fontsize{8.0}{9.5}\selectfont] at (376.5,203) {$s_i^{k''}$};
  \node[labelbox,text=pathblue,font=\bfseries\fontsize{8.0}{9.5}\selectfont] at (567.5,57) {$s_i^{M_i}$};
  \node[labelbox,text=currentred,font=\bfseries\fontsize{8.5}{10}\selectfont] at (320,369) {$\mathcal F_i^{k'}$};
  \node[labelbox,text=futureorange,font=\bfseries\fontsize{8.0}{9.5}\selectfont] at (491.5,203) {$\mathcal F_i^{k''}$};


  \draw[textdark,line width=.85pt,-{Triangle[length=5pt,width=6pt]}] (60,460)--(115,460);
  \draw[textdark,line width=.85pt,-{Triangle[length=5pt,width=6pt]}] (60,460)--(60,405);
  \node[legendtext,anchor=west] at (115,460) {$x$};
  \node[legendtext,anchor=south] at (60,405) {$y$};



\end{tikzpicture}%
    }
    \caption{%
        Discrete spatio-temporal representation for vehicle $i$'s path. The reference positions $s_i^{k'}$ and $s_i^{k''}$ illustrate two discrete positions along the path, with corresponding vehicle footprints $\mathcal{F}_i^{k'}$ and $\mathcal{F}_i^{k''}$. Each footprint is associated with its planned time $t_{k'}$ or $t_{k''}$ and forms part of
the reservation request $\mathcal{Q}_i =  \left\{(\mathcal{F}_i^k,t_k)\right\}_{k=0}^{M_i}$.
    }
    \label{fig:velocity_spatial_discretization}
\end{figure}
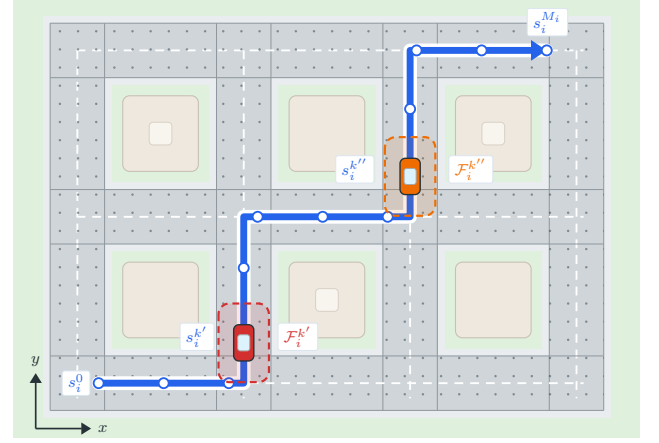

\subsection{SPATIO-TEMPORAL DISCRETE MODEL}
\label{sec:discretization}

VeloCity solves the optimization problem in the mixed-integer domain. Therefore, it relies on a discrete space-time representation of the urban area. Given the predefined
path $\mathcal{P}_i$, each vehicle trajectory is discretized with spatial
resolution $\Delta s$.
Accordingly, the coordinate $s_i(t)$ is mapped into a
sequence of $M_i$ spatial cells:
\begin{equation}
    \mathcal{S}_i =
    \left\{
    s_i^0, s_i^1,\ldots,s_i^{M_i}
    \right\},
\end{equation}
where consecutive elements satisfy
$s_i^{k+1}-s_i^k = \Delta s$.
Each spatial cell is associated with a two-dimensional position through the
trajectory mapping $\gamma_i(\cdot)$.
To preserve physical feasibility, each vehicle is represented as a sequence
of occupied space-time regions. Let
$\mathcal{F}_i^k$ denote the set of spatial cells occupied by vehicle $i$
when its reference position is $s_i^k$. The corresponding reservation request
is therefore:
\begin{equation}
    \mathcal{Q}_i =
    \left\{
    (\mathcal{F}_i^k,t_k)
    \right\}_{k=0}^{M_i}.
\end{equation}
This representation allows collision checking to be converted into a discrete
space-time conflict test:
\begin{equation}
    \mathcal{F}_i^k
    \cap
    \mathcal{R}(t_k)
    =
    \emptyset ,
    \quad
     k = 0, \ldots, M_i,
\label{eq:discrete_collision}
\end{equation}
where $\mathcal{R}(t)$ is the current reservation table maintained by the
coordinator. A representation of a vehicle trajectory is provided in Fig.~\ref{fig:velocity_spatial_discretization}. 

\subsection{DECENTRALIZED MOBILITY OPTIMIZATION}
\label{sec:distributed_optimization}

Instead of solving the joint optimization problem in
\eqref{eq:objective}, {VeloCity} decomposes the problem into individual
vehicle optimizations. 
When vehicle $i$ enters the coordination area, it retrieves the current
reservation table $\mathcal{R}$ and locally solves:
\begin{equation}
\begin{aligned}
\min_{\mathbf{a}_i}
\quad&
t_{\mathrm{out},i}-t_{\mathrm{in},i}
\\
\text{s.t.}\quad
&
a_{\min,i}\leq a_i(t)\leq a_{\max,i}\,,
\\
&
0\leq v_i(t)\leq
\min(V_{\max,i},V_{\max}(\gamma_i(s_i(t))))\,,
\\
&
\mathcal{F}_i(t)
\cap
\mathcal{R}(t)
=
\emptyset \,, 
\\
&
\forall t \geq t_{\mathrm{in}, i}\,.
\end{aligned}
\label{eq:local_problem}
\end{equation}
The local problem preserves the constraints of the original MINLP while
removing the coupling among all vehicles. Consequently, each CAV only needs
to consider previously committed trajectories rather than the complete fleet
state. 
%
%
%
%
The local optimization in \eqref{eq:local_problem} is solved through a
best-first search operating on the constructed state space.
The search graph is defined as a labeled directed graph:
\begin{equation}
    \mathcal{G}_i=(\mathcal{X}_i,\mathcal{E}_i),
\end{equation}
where each node $\mathbf{x}_i(t) \in \mathcal{X}_i$ represents a feasible state. 
The state of vehicle $i$ is defined as:
\begin{equation}
    \mathbf{x}_i(t)=
    \left(s_i(t),v_i(t),t\right),
\end{equation}
where the discrete speed evolution follows:
\begin{equation}
    v_i(t+\Delta t)
    =
    v_i(t)+a_i(t)\Delta t ,
\label{eq:velocity_update}
\end{equation}
and the position evolves according to:
\begin{equation}
    s_i(t+\Delta t)
    =
    s_i(t)+v_i(t+\Delta t)\Delta t .
\label{eq:position_update}
\end{equation} 
Each edge is a directed triple:
\begin{equation}
    \left(\mathbf{x}_i(t), a_i(t), \mathbf{x}_i(t + \Delta t)\right) \in \mathcal{E}_i ,
\end{equation}
which represents the transition from the current state $\mathbf{x}_i(t)$ to the successor state $\mathbf{x}_i(t + \Delta t)$ driven by the specific admissible acceleration command $a_i(t)$. The set of admissible successors is further constrained by the reservation
table. A transition is
valid only if:
\begin{equation}
\label{eq:reservation_check}
    \mathcal{F}_i(\mathbf{x}_i(t+\Delta t))
    \cap
    \mathcal{R}(t+\Delta t)
    =
    \emptyset .
\end{equation}
This condition prevents conflicts with previously committed vehicles.
The optimal solution is obtained by solving a shortest-path problem on
$\mathcal{G}_i$. Since the objective is to
minimize the traversal time, the transition cost is defined as the elapsed
time:
\begin{equation}
    c(
    \mathbf{x}_i(t),
    \mathbf{x}_i(t+\Delta t)
    )
    =
    \Delta t\,.
\end{equation}
Consequently, the accumulated cost of a path corresponds exactly to the
arrival time:
\begin{equation}
    g(\mathbf{x}_i(t))
    = \sum_{\tau=t_{\mathrm{in},i}}^{t}
    c(
    \mathbf{x}_i(\tau),
    \mathbf{x}_i(\tau+\Delta t)
    ) =
    (t-t_{\mathrm{in},i}) \Delta t\,.
\end{equation}
To determine the shortest-path, each candidate state $\mathbf{x}_i(t)$ is evaluated by:
\begin{equation}
    f(\mathbf{x}_i(t))
    =
    g(\mathbf{x}_i(t))+h(\mathbf{x}_i(t)),
\end{equation}
where the heuristic term estimates the remaining travel time:
\begin{equation}
\label{eq:astar_heuristic}
    h(\mathbf{x}_i(t))
    =
    \frac{D_i-s_i}{V_{\max,i}} .
\end{equation}
The heuristic in \eqref{eq:astar_heuristic} is admissible because it assumes
that the vehicle can immediately travel at its maximum allowed speed, thus
providing a lower bound on the remaining traversal time.
Once the goal condition
\begin{equation}
    s_i(t)\geq D_i
\end{equation}
is reached, the resulting sequence of states directly determines the vehicle
mobility profile:
\begin{equation}
    \Pi_i
    =
    \left\{
    (s_i(t),v_i(t),a_i(t))
    \right\}_{t=t_{\mathrm{in},i}}^{t_{\mathrm{out},i}} .
\end{equation}
The computed profile is then converted into a sequence of physical
space-time reservations $\mathcal{Q}_i$, which are submitted to the coordinator. 
Upon reception, the coordinator performs:
\begin{equation}
    \mathcal{R}
    \leftarrow
    \mathcal{R}
    \cup
    \mathcal{Q}_i .
\end{equation}
Therefore, future vehicles automatically avoid already committed trajectories.
The coordinator only performs lightweight consistency checking, while the
computational effort remains distributed among the CAVs.

\subsection{COMPUTATIONAL COMPLEXITY ANALYSIS}
\label{sec:complexity}

By decomposing the MINLP in \eqref{eq:objective} into independent local searches, the computational complexity of the framework grows approximately linearly with the number of vehicles, as each \ac{CAV} solves only one sequential best-first search. For a given vehicle $i$, the size of the local search graph $\mathcal{G}_i=(\mathcal{X}_i,\mathcal{E}_i)$ introduced in Section~\ref{sec:methodology}.\ref{sec:distributed_optimization} is governed by three primary parameters:
\begin{itemize}
    \item $M_i = D_i/\Delta s$, the total number of discrete spatial cells along the predefined path $\mathcal{P}_i$;
    \item $V_i = V_{\max,i}/\Delta v$, the number of distinct speed levels induced by the discrete dynamics in \eqref{eq:velocity_update};
    \item $b=|\{a_{\min,i},\dots,a_{\max,i}\}|$, the branching factor, representing the cardinality of the admissible discrete acceleration set.
\end{itemize}
Because vehicles cannot travel backward ($v_i(t) \ge 0$), and because the objective function strictly minimizes the arrival time, the algorithm is constrained to move forward through the spatial domain. Consequently, the maximum number of states the algorithm can expand is bounded by the combinations of spatial positions and speed levels:
\begin{equation}
    |\mathcal{X}_i| \;\le\; O(M_i \, V_i).
    \label{eq:state_space_bound}
\end{equation}

\subsubsection{Per-Vehicle Search Complexity}

To determine the optimal mobility profile $\Pi_i$, Algorithm~\ref{alg:astar} employs a best-first search utilizing a binary-heap open list and a hash-indexed closed list. Considering the reservation table $\mathcal{R}$ stored as a hash set indexed by discretized space-time cells, the worst-case time complexity for a single vehicle is:
\begin{equation}
    O\!\left(b\,|\mathcal{X}_i|\log|\mathcal{X}_i|\right)
    =
    O\!\left(b\,M_iV_i\log(M_iV_i)\right),
    \label{eq:per_vehicle_complexity}
\end{equation}
with an auxiliary memory footprint bounded by $O(|\mathcal{X}_i|) = O(M_iV_i)$. 
This efficiency arises because the admissible heuristic defined in \eqref{eq:astar_heuristic} guarantees that the search expands each reachable state in $\mathcal{X}_i$ at most once, resulting in at most $M_iV_i$ expansions. During each expansion, the algorithm evaluates up to $b$ successors. For each successor, it executes a constant-time $O(1)$ kinematic propagation and a $O(1)$ conflict test against $\mathcal{R}$ via \eqref{eq:reservation_check}. Consequently, the dominant operation per state is the insertion or key-update within the priority queue, scaling at $O(\log(M_iV_i))$ for a binary-heap open list. 
Following the search, the reservation commit step ($\mathcal{R}\leftarrow\mathcal{R}\cup\mathcal{Q}_i$) inserts the sequence of space-time footprints into the global hash set. Because the path length is fundamentally limited by the number of spatial cells, this step requires at most $O(M_i)$, which is dominated by \eqref{eq:per_vehicle_complexity}.

\subsubsection{System-Level Complexity}

At the system level, Algorithm~\ref{alg:velocity_main} processes incoming traffic sequentially. Each \ac{CAV} solves its local problem against the reservations committed by its predecessors. For a total fleet of $N$ vehicles, the theoretical overall system running time scales as:
\begin{equation}
    O\!\left(
    \sum_{i=1}^{N} b\,M_iV_i\log(M_iV_i)
    \right)
    =
    O\!\left(N\,b\,\bar{M}\bar{V}
    \log(\bar{M}\bar{V})\right),
    \label{eq:system_complexity}
\end{equation}
where $\bar{M}$ and $\bar{V}$ represent upper bounds across the heterogeneous fleet. While \eqref{eq:system_complexity} demonstrates that the global complexity is strictly \emph{linear} with respect to $N$, this theoretical worst-case bound severely overestimates the practical computational burden in realistic urban scenarios. Indeed, in these areas, vehicles do not accumulate indefinitely. As \acp{CAV} reach their destinations and exit the coordination area, their spatio-temporal footprints are left in the past. Because the reservation table intrinsically encodes time, a vehicle entering the network at $t_{\mathrm{in},i}$ only evaluates conflict checks against the reservations of currently active vehicles whose temporal horizons overlap with its own. Consequently, the local planning process is completely independent of the historical fleet. This natural spatio-temporal decoupling ensures that the operational lookup time against $\mathcal{R}$ remains constant and bounded by the instantaneous traffic density, rather than growing monotonically with the cumulative vehicle count $N$.

\begin{figure*}[!t]
    \centering
    \subfloat[Tokyo]{\includegraphics[width=0.24\textwidth]{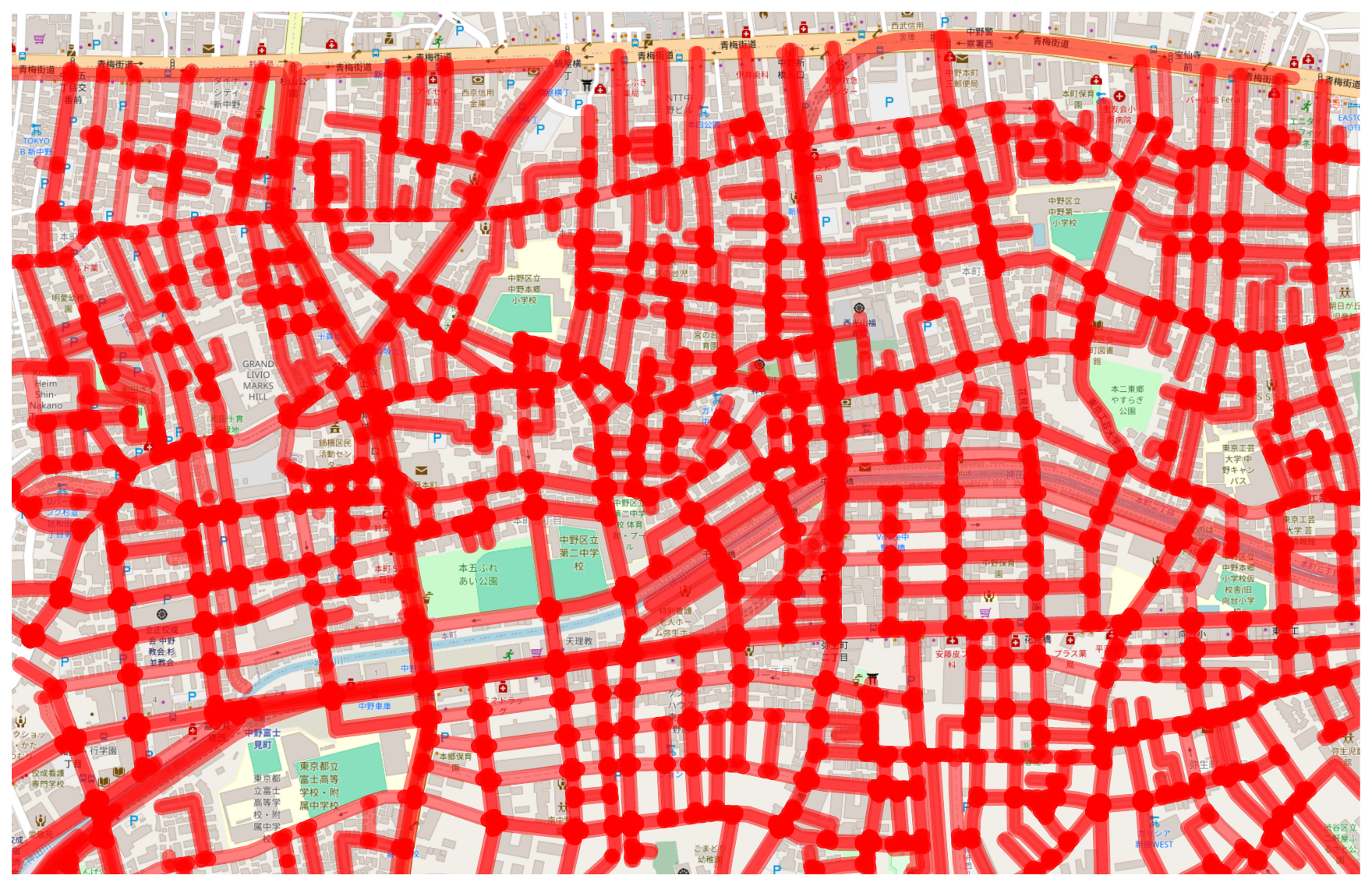}\label{fig:map_tokyo}}
    \hfil
    \subfloat[Manhattan]{\includegraphics[width=0.24\textwidth]{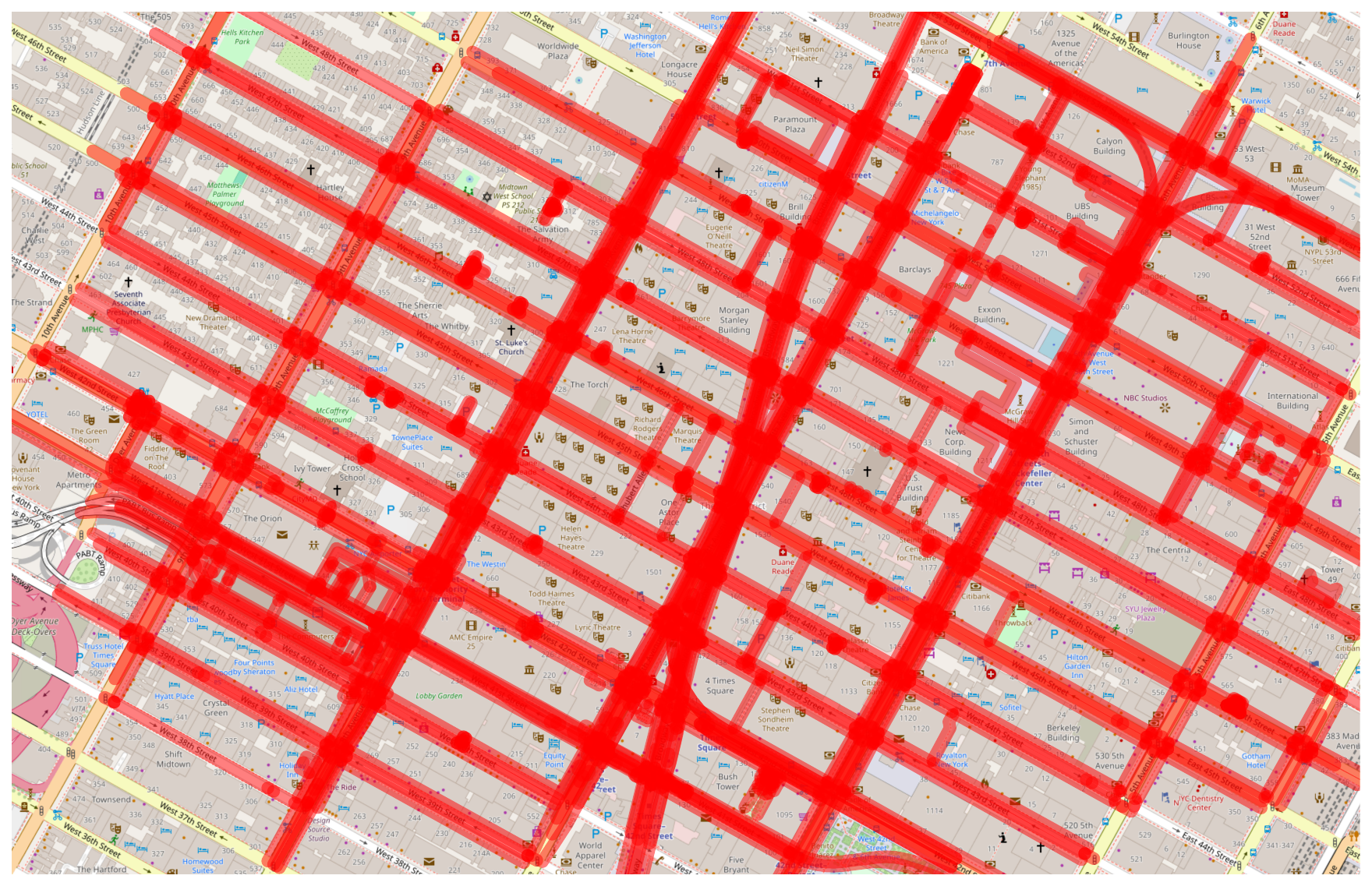}\label{fig:map_manhattan}}
    \hfil
    \subfloat[Rome]{\includegraphics[width=0.24\textwidth]{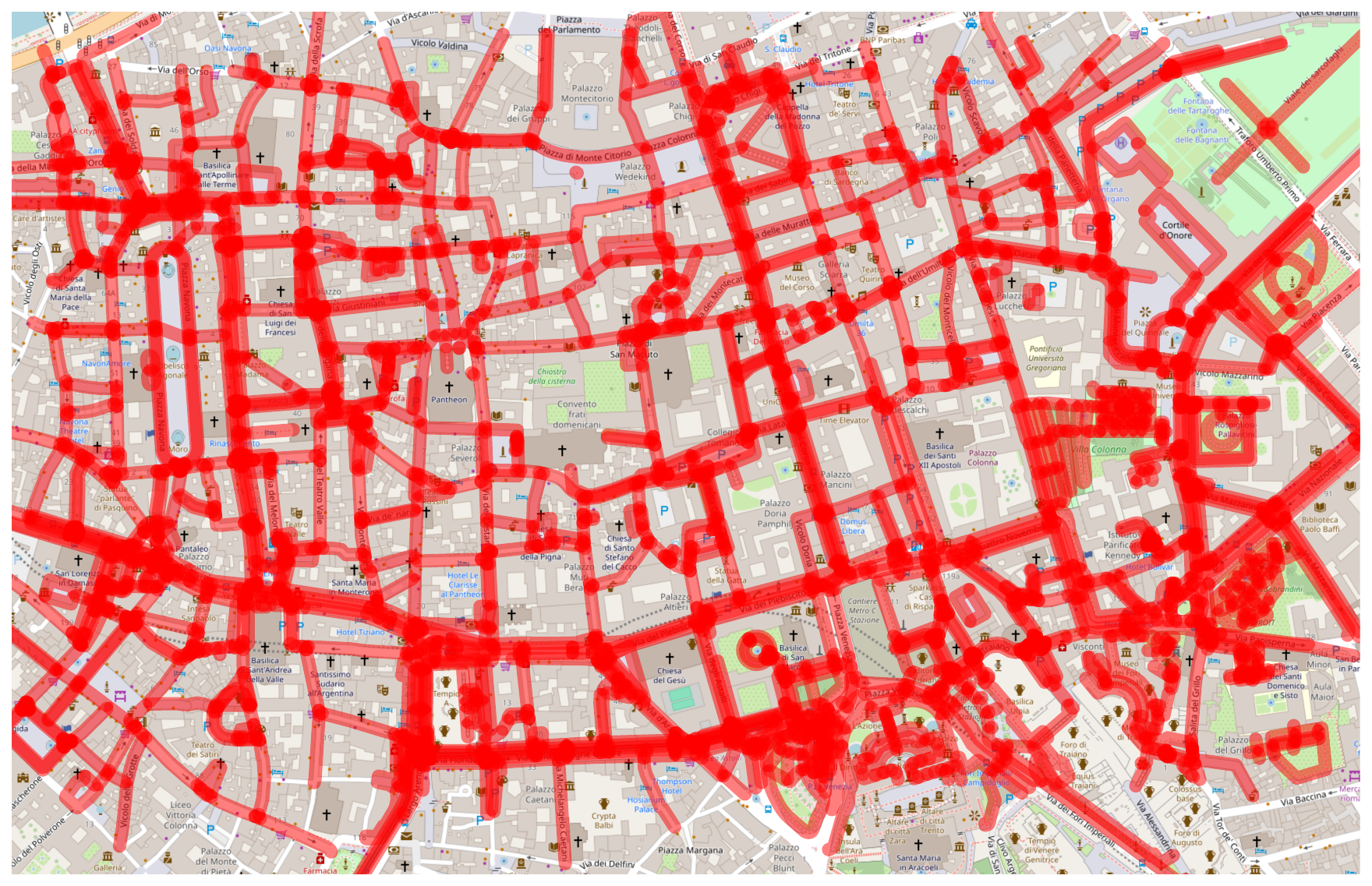}\label{fig:map_rome}}
    \hfil
    \subfloat[Bologna]{\includegraphics[width=0.24\textwidth]{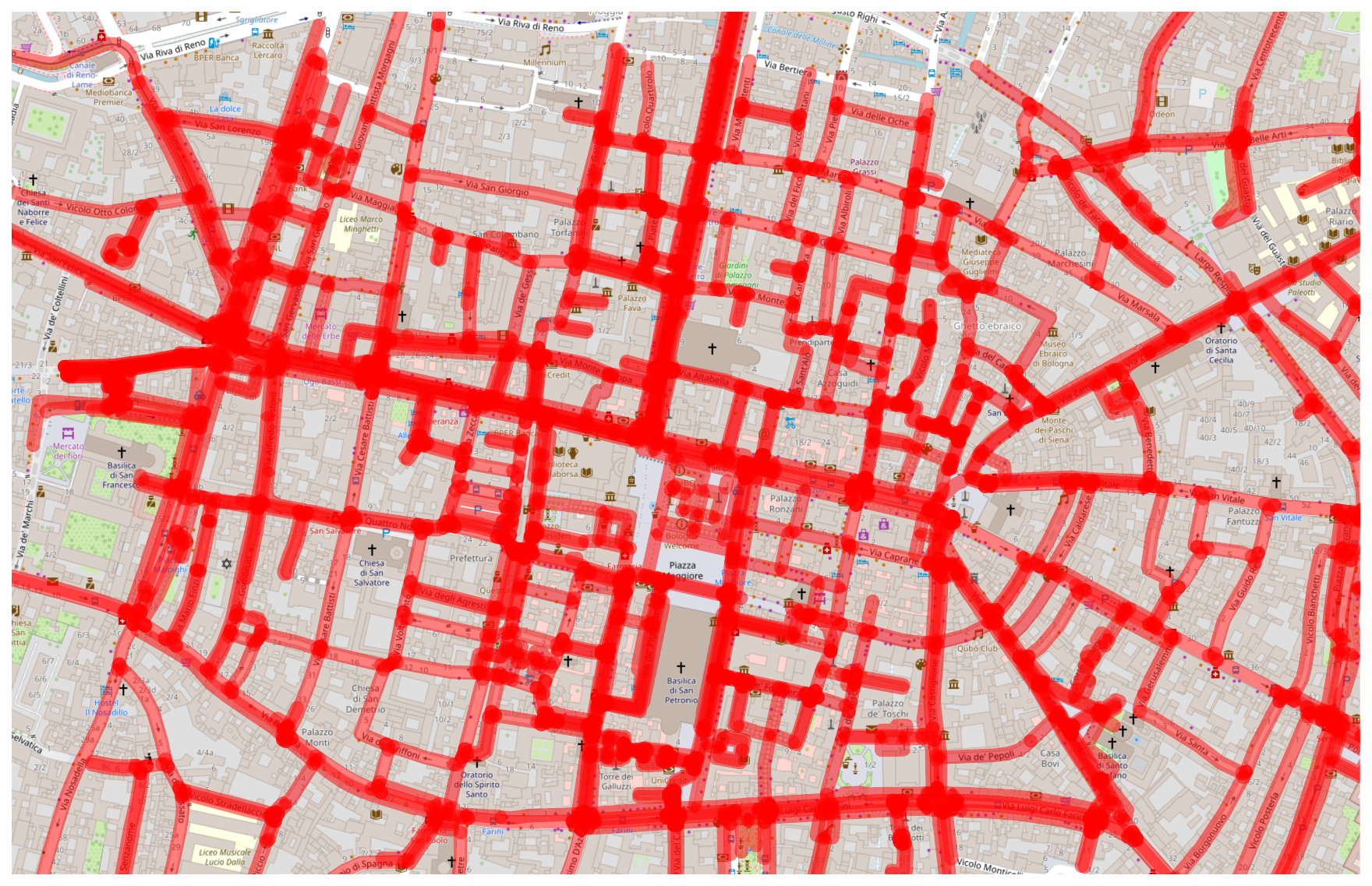}\label{fig:map_bologna}}
    \caption{Experimental OpenStreetMap areas for the four evaluated urban scenarios.}
    \label{fig:maps}
\end{figure*}

\begin{figure*}[!t]
    \centering
    \subfloat[Tokyo]{\includegraphics[width=0.24\textwidth]{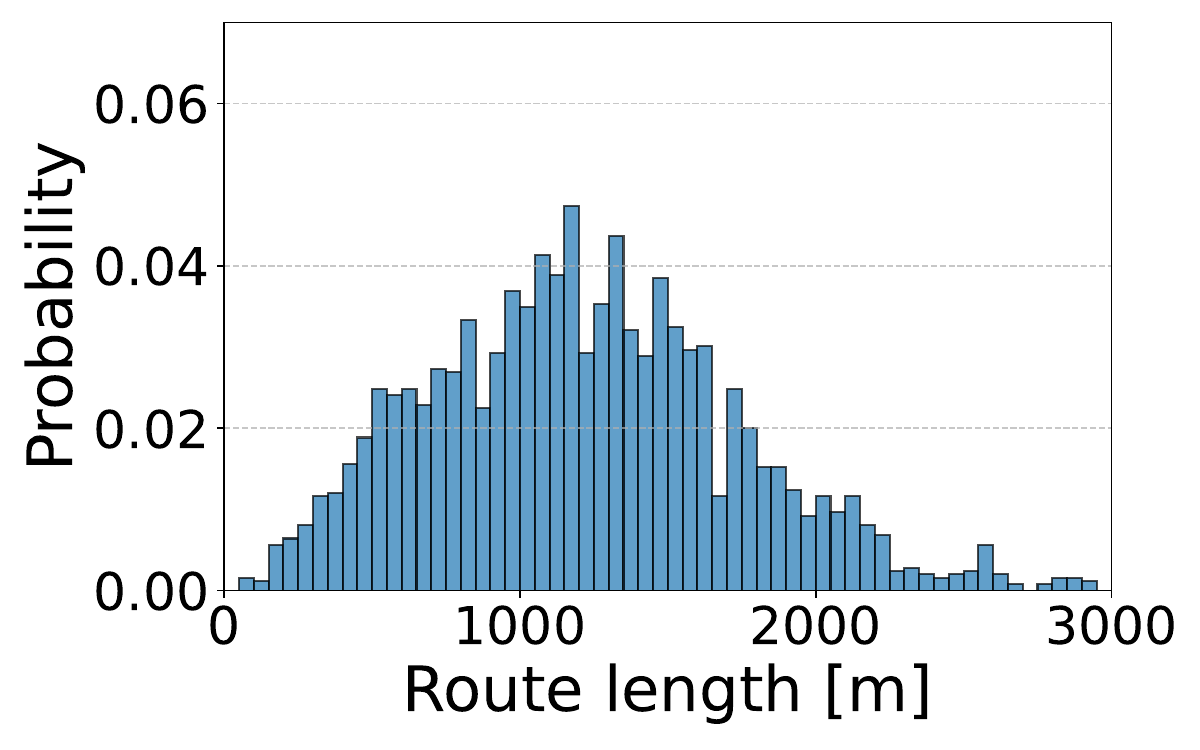}\label{fig:pmf_tokyo}}
    \hfil
    \subfloat[Manhattan]{\includegraphics[width=0.24\textwidth]{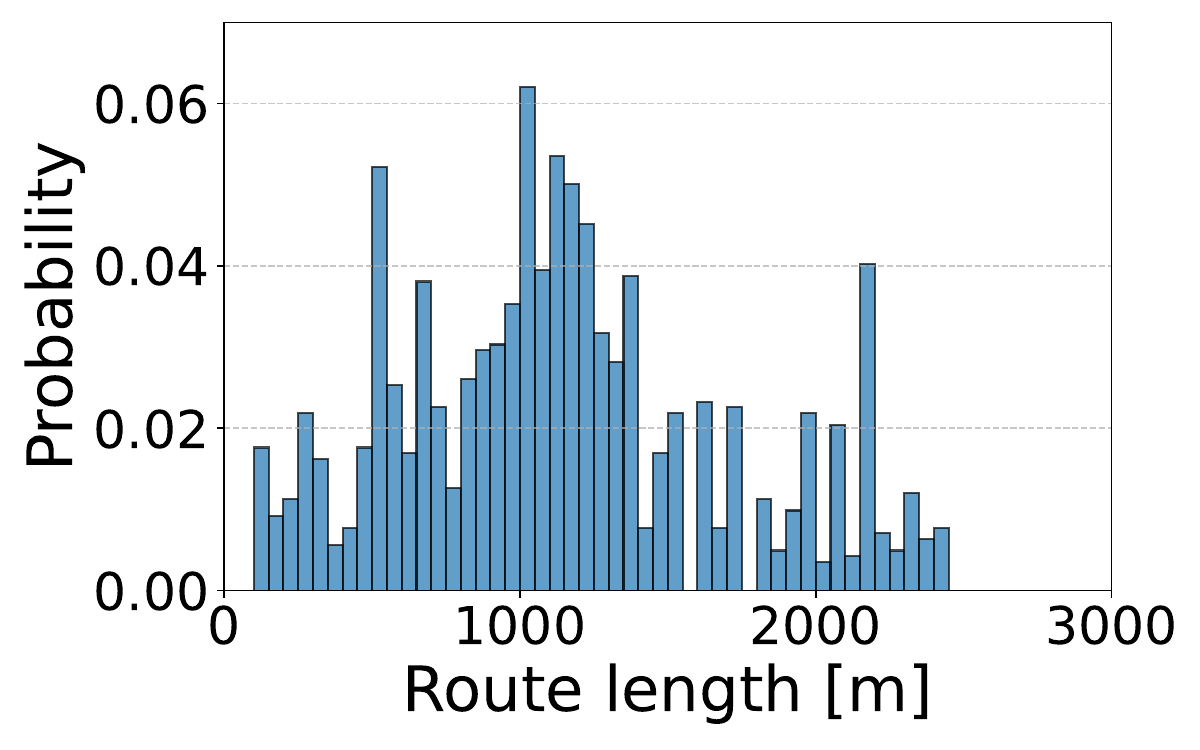}\label{fig:pmf_manhattan}}
    \hfil
    \subfloat[Rome]{\includegraphics[width=0.24\textwidth]{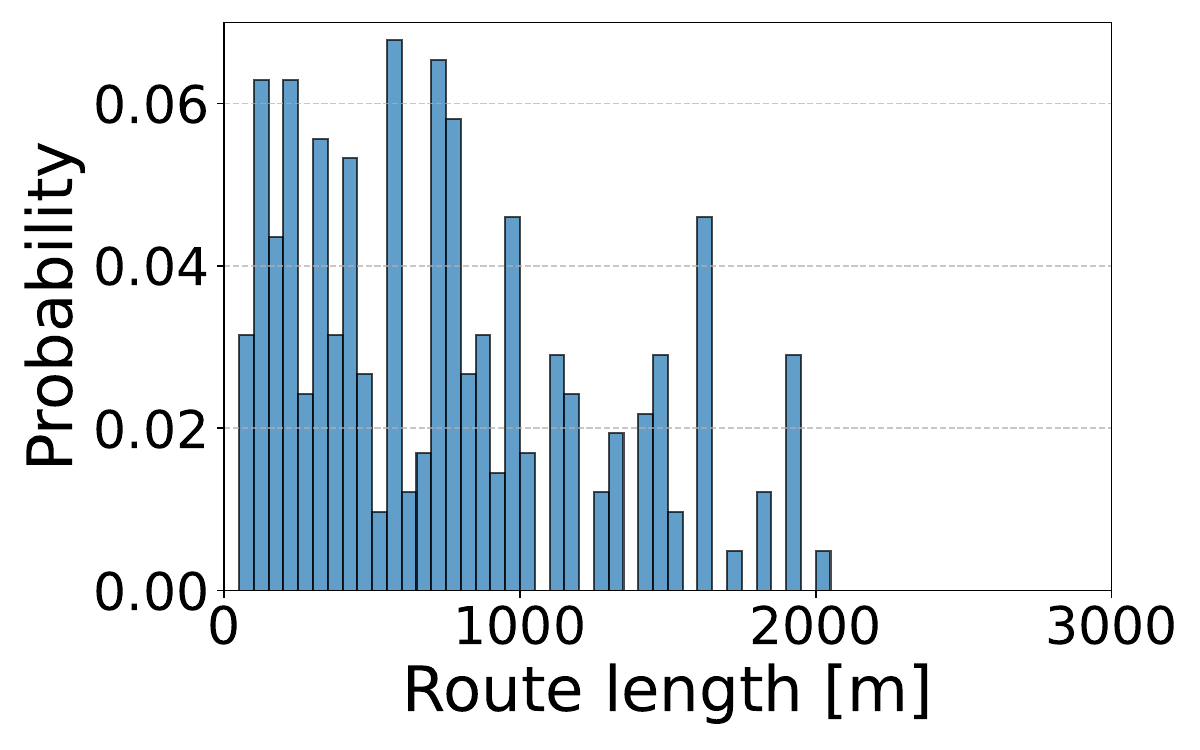}\label{fig:pmf_rome}}
    \hfil
    \subfloat[Bologna]{\includegraphics[width=0.24\textwidth]{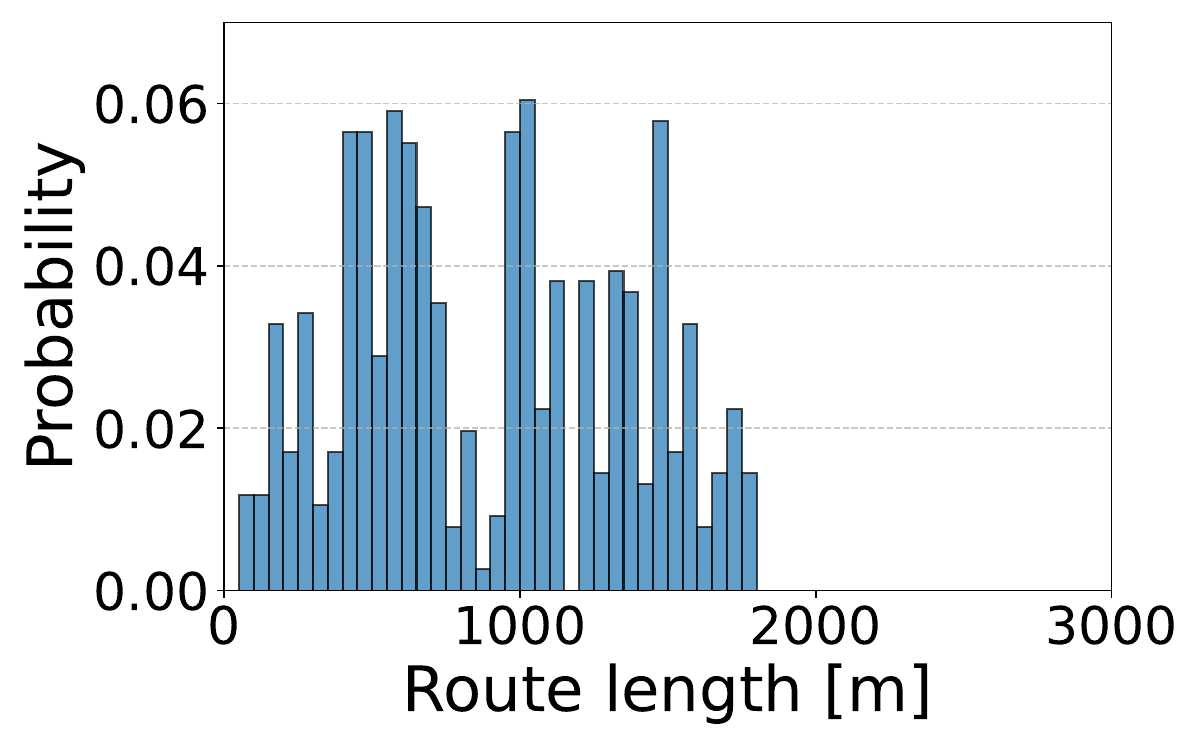}\label{fig:pmf_bologna}}
    \caption{Empirical \ac{PMF} of vehicle route lengths for the four evaluated urban scenarios.}
    \label{fig:pmfs}
\end{figure*}

\begin{table*}[htbp]
    \centering
    \caption{Spatial metrics of the evaluated urban networks}
    \label{tab:spatial_metrics}
    
    \begin{tabular}{lccc}
        \hline
        \textbf{Scenario} & \textbf{Area} & \textbf{Road Length} & \textbf{Bounding Box ($\text{Lon} \times \text{Lat}$)} \\
        \hline
        Tokyo     & $3.07$ $\mathrm{km}^2$ & $93.68$ $\mathrm{km}$ & $[139.66, 139.68] \times [35.68, 35.69]$ \\
        Manhattan & $4.24$ $\mathrm{km}^2$ & $102.98$ $\mathrm{km}$ & $[-73.99, -73.97] \times [40.75, 40.76]$ \\
        Rome      & $1.73$ $\mathrm{km}^2$ & $60.88$ $\mathrm{km}$ & $[12.47, 12.48] \times [41.89, 41.90]$ \\
        Bologna   & $2.54$ $\mathrm{km}^2$ & $57.62$ $\mathrm{km}$ & $[11.33, 11.35] \times [44.49, 44.49]$ \\
        \hline
    \end{tabular}
\end{table*}

\subsection{KEY-PROPERTIES OF VELOCITY}
\label{sec:properties}

The architectural choices underlying the {VeloCity} framework yield several advantageous properties for urban traffic coordination:

\begin{enumerate}

\item The framework resolves the highly complex global coordination problem through \emph{sequential} and \emph{decentralized optimizations}. As each vehicle queries the coordinator, computes its local mobility profile, and commits its reservation, the system is immediately ready for the next incoming agent. This approach is highly realistic and computationally practical for real-time edge-computing deployment, avoiding costly joint \ac{CAV} fleet optimization.

\item The reservation-based representation is \emph{independent of any specific road topology}. Because collision avoidance is evaluated directly in physical space-time coordinates, the same mathematical formulation natively supports arbitrary urban layouts, including complex intersections, roundabouts, lane merges, and multi-leg junctions, without requiring custom conflict-resolution rules.

\item The explicit integration of vehicle kinematics into the search space guarantees that the generated \emph{trajectories are not only collision-free but also physically executable}. 
Furthermore, the geometric footprint $\mathcal{F}_i$ offers additional flexibility; it can be enlarged to reserve extra spatial margins, effectively buffering against uncertainties related to low-level control tracking errors or physical actuation delays.

\item Driven by the objective to minimize arrival time without imposing a hard deadline, the framework gains flexibility. A vehicle can always find a feasible mobility profile by opting to enter the managed area, decelerate, and wait indefinitely until downstream congestion clears. Consequently, a planning fault is triggered \emph{if and only if} the vehicle lacks sufficient space even to safely enter and immediately halt. \emph{Unavoidable deadlocks are thus detected instantaneously at the boundaries}, preventing cascading gridlocks within the core of the infrastructure.

\item In severely congested scenarios where an immediate initial access request fails, the framework is easily extended to support relaxing the requested entry time $t_{\mathrm{in},i}$. \emph{By allowing a vehicle to delay its entry, an optimal and collision-free crossing plan can always be identified}. This temporal relaxation is particularly suited for vehicles departing from parking spots or those waiting in structured access queues at the control area's perimeter.

\item To accommodate heterogeneous urban areas, \emph{the coordinator can pre-initialize static or periodic reservations} within $\mathcal{R}$ prior to scheduling incoming \acp{CAV}. Let $\mathcal{R}_{\mathrm{prior}}(t)$ represent the spatio-temporal space pre-allocated to unmanaged traffic entities. This includes public transportation assets (e.g., scheduled tram or bus), deterministic pedestrian crossing windows, or any pre-configured users. Consequently, any newly entering \ac{CAV} must find a mobility profile that is mutually compliant with both existing vehicular allocations and these background infrastructure constraints, satisfying:
\begin{equation}
\label{eq:const_no_overlap}
    \mathcal{F}_i(s_i(t)) \cap \left( \mathcal{R}_{\mathrm{veh}}(t) \cup \mathcal{R}_{\mathrm{prior}}(t) \right) = \emptyset\,,
\end{equation}
$\forall t \geq t_{\mathrm{in}, i}$, where $\mathcal{R}_{\mathrm{veh}}(t)$ denotes the union of footprints from previously committed \acp{CAV}.

\item Unlike distributed consensus algorithms that require continuous \ac{V2V} state sharing, VeloCity \emph{preserves privacy} and \emph{minimizes network overhead}. Vehicles only exchange once their intended spatio-temporal footprints with the coordinator via \ac{V2I} links, without exposing exact destinations, internal control parameters, or complete routing intentions to the broader fleet.


\item By processing reservation requests in the order they arrive at the coordinator, the framework enforces a fair sequential allocation of space-time slots. This guarantees deterministic access, ensuring that early vehicles are not starved by subsequent high-density traffic.

\end{enumerate}

\section{NUMERICAL RESULTS}
\label{sec:experiments}

In this section, we extensively evaluate the performance of VeloCity across four large-scale real-world urban scenarios.

\subsection{EXPERIMENTAL SETUP}

To evaluate the performance and scalability of the proposed framework, we conduct extensive simulations across four distinct and highly heterogeneous urban areas: Tokyo, Manhattan, Rome, and Bologna. The geographic road networks for these scenarios are illustrated in Fig.~\ref{fig:maps}. To ensure a high degree of realism in our mobility modeling, the network topology for each scenario is imported directly from OpenStreetMap, including lane definitions, explicit restrictions at intersections, legal speed limits, and detailed intersection layouts.
The selected areas offer diverse topological characteristics, ranging from highly structured grid layouts to irregular historical streets. To quantify this diversity, the specific spatial metrics for each evaluated scenario are summarized in Table~\ref{tab:spatial_metrics}.

All numerical experiments simulate traffic operations over a duration of one hour in \ac{SUMO}~\cite{sumoRef}. Realistic traffic demand is generated utilizing the \texttt{randomTrips.py} utility provided by the \ac{SUMO} simulator. This tool automatically generates randomized but physically valid routes across the imported OpenStreetMap networks, ensuring that vehicles are spawned at valid distributed edges and assigned achievable destinations within the boundaries of the control area. To characterize the resulting traffic demand, Fig.~\ref{fig:pmfs} reports the empirical \ac{PMF} of the vehicle route lengths for each of the four urban scenarios. 

\begin{figure*}[t]
    \centering

    \includegraphics[width=\textwidth]{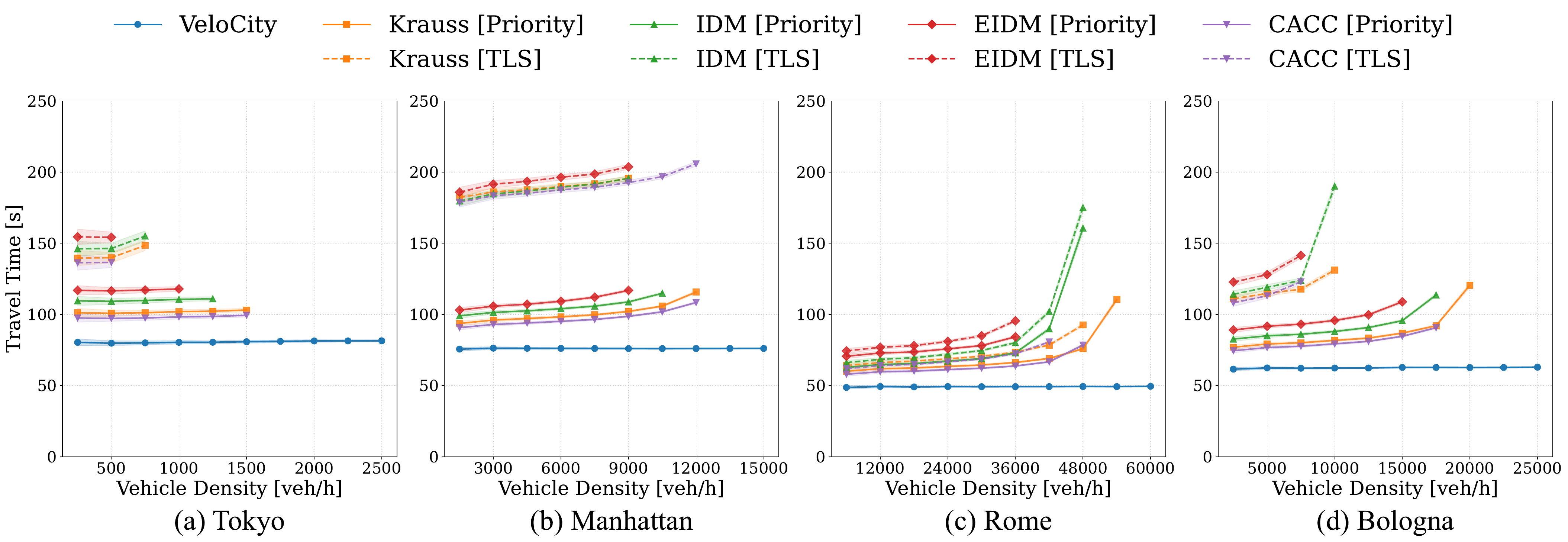}
    \caption{Vehicle travel time for the four evaluated urban scenarios.}
    \label{fig:travel_time}
\end{figure*}

\begin{figure}[t]
    \centering

    \includegraphics[trim={12cm 2cm 9cm 2cm}, clip,width=0.7\columnwidth]{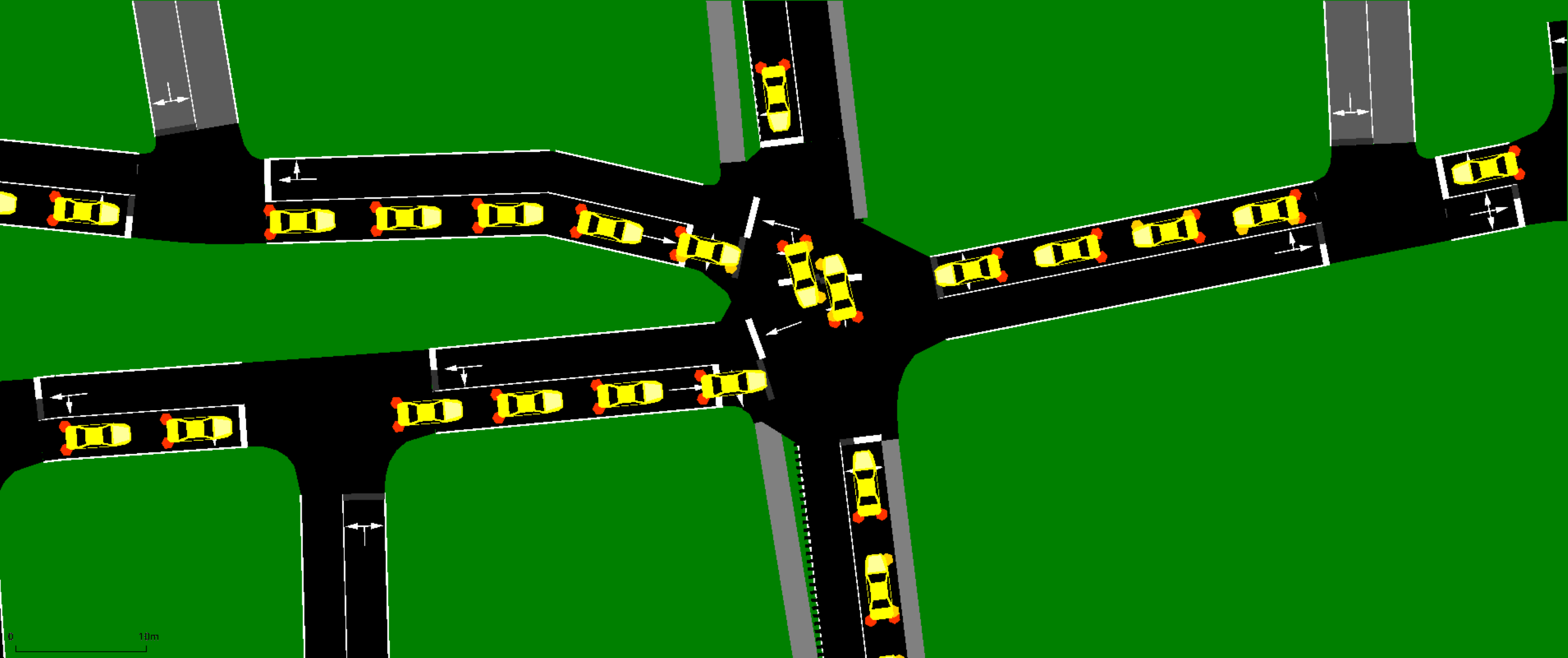}
    \caption{Example of a congestion event in the Tokyo scenario incurred by baselines.}
    \label{fig:congestion}
\end{figure}

Within \ac{SUMO}, the discrete space-time model operates with a temporal resolution of $\Delta t = 0.5$\,s and a spatial resolution of $\Delta s = 0.5$\,m. To reflect typical urban driving behavior, all \acp{CAV} feature a maximum speed equal to $V_{\max, i} = 50$\,km/h. Acceleration is bounded by a maximum capacity of $a_{\max, i} = 1$\,m/s$^2$ and a maximum deceleration of $a_{\min, i} = -2$\,m/s$^2$. All generated vehicles are 5\,m long, 2\,m wide. Furthermore, to ensure robust collision avoidance and to account for physical actuation uncertainties (as outlined in the third item in Section~\ref{sec:properties}), the geometric footprint $\mathcal{F}_i$ is defined to incorporate a strict safety margin. Consequently, an area of $2.5$\,m around each vehicle's boundaries is reserved during the spatio-temporal allocation process.

\subsection{BENCHMARKS}
\label{sec:benchmarks}

To evaluate the performance of the proposed VeloCity framework, we compare it against a set of established car-following models natively implemented within the SUMO. Specifically, we consider the following models:
\begin{enumerate}
    \item \textit{Krauss:} The default car-following model in SUMO. It computes a safe speed based on the leading vehicle's speed and the available gap, incorporating a stochastic imperfection parameter to simulate realistic human driving behavior and reaction delays~\cite{krauss1998microscopic}.
    \item \textit{\ac{IDM}:} A continuous-time car-following model that transitions between free-road acceleration and collision-avoidance deceleration. It is governed by a desired target speed and a safe time-headway parameter~\cite{treiber2000congested}.
    \item \textit{\ac{EIDM}:} An enhancement of the classical \ac{IDM} specifically calibrated for sub-second simulation resolutions. The \ac{EIDM} introduces explicit reaction times, driver estimation errors, and jerk limitations to yield more accurate acceleration profiles in dense urban traffic~\cite{salles2020extending}.
    \item \textit{\ac{CACC}:} A deterministic model representing connected autonomous vehicles that leverage \ac{V2V} communications. By sharing acceleration and speed data, \ac{CACC} vehicles can safely maintain tight inter-vehicle headways and string stability~\cite{milanes2014modeling}.
\end{enumerate}
The evaluation of these baseline models is split into two distinct operational settings across the simulated networks:
\begin{enumerate}
    \item \textit{Signalized Control:} Intersections within the managed urban areas are possibly governed by \ac{TLS}, according to their OpenStreetMap data. Vehicles operating under the baseline car-following models must react to cyclic red/green phases in addition to their leading vehicles.
    \item \textit{Autonomous Priority Control:} Traffic lights are completely disabled. Intersections operate based on right-of-way priority rules (e.g., yield and stop mechanics). Vehicles relying on the benchmark models must dynamically identify safe spatial gaps in the crossing traffic streams before proceeding.
\end{enumerate}
Conversely, VeloCity inherently overrides both traditional traffic lights and static priority rules. Vehicles operating under the proposed framework navigate the identical un-signalized network layouts by relying entirely on the decentralized spatio-temporal algorithm introduced in Section~\ref{sec:methodology}. Overall, our experimental analysis compares the proposed methodology with eight alternative approaches. 

\begin{figure*}[t]
    \centering

    \includegraphics[width=\textwidth]{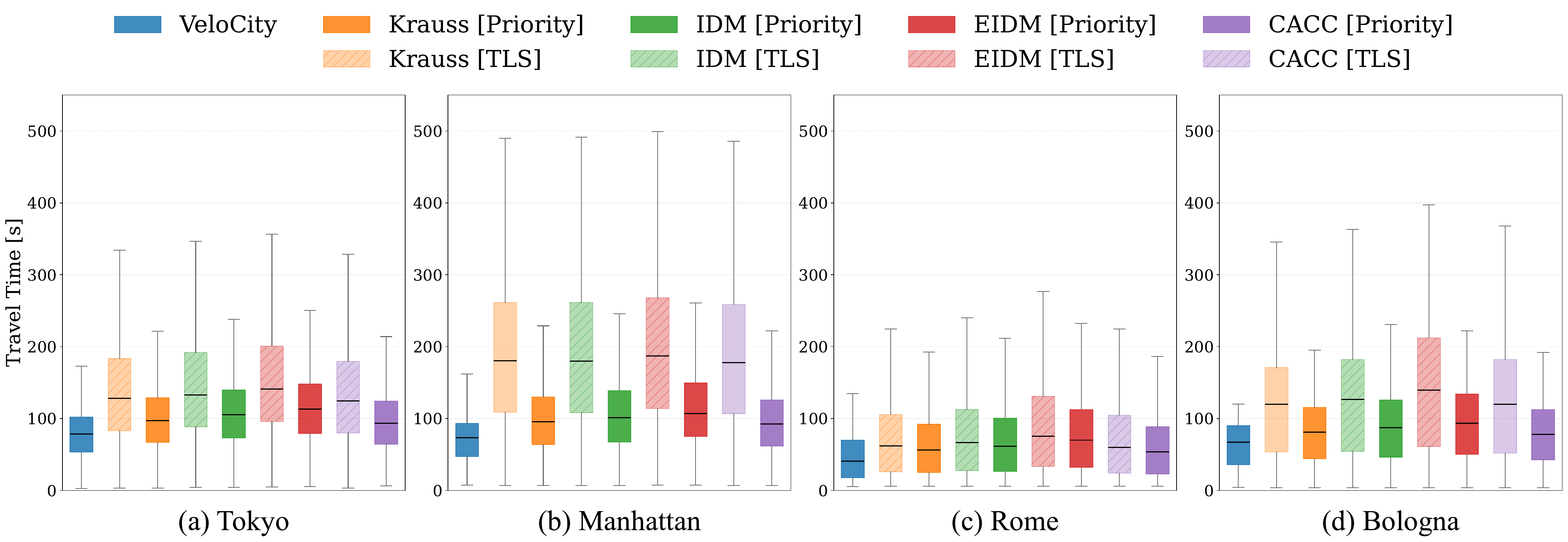}
    \caption{Vehicle travel time distribution for the four evaluated urban scenarios, comprising (a) 500 veh/h for Tokyo, (b) 9,000 veh/h for Manhattan, (c) 36,000 veh/h for Rome, and (d) 7,500 veh/h for Bologna.}
    \label{fig:box_plot}
\end{figure*}

\begin{figure*}[t]
    \centering

    \includegraphics[width=\textwidth]{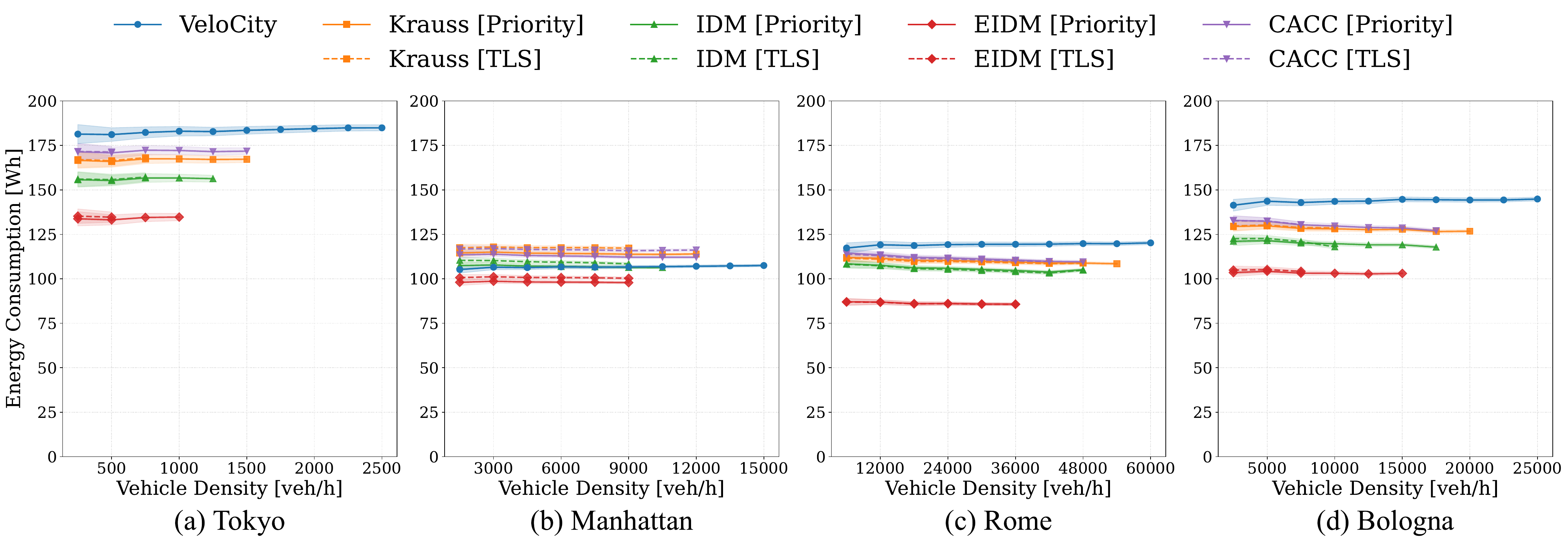}
    \caption{Vehicle energy consumption for the four evaluated urban scenarios.}
    \label{fig:energy}
\end{figure*}

\subsection{TRAVEL TIME AND ENERGY CONSUMPTION ANALYSIS}





To comprehensively evaluate the proposed framework, we assess both the travel time and energy consumption metrics across the CAV fleets. For the energy analysis, we assume a fleet composed entirely of fully electric vehicles. The experiments are conducted by varying the traffic demand, defined in terms of vehicles per hour. To ensure statistical robustness, each scenario-density combination is evaluated over five independent simulation runs with varying random seeds.

The aggregate travel time results are depicted in Fig.~\ref{fig:travel_time}, with data reported in terms of average per-vehicle travel time and 95\% confidence intervals. Across all four urban areas, VeloCity demonstrates superior efficiency and remarkable stability as density increases. Indeed, the average delay for vehicles managed by VeloCity remains practically constant across the entire tested density spectrum for each city. For example, in the Tokyo scenario, VeloCity maintains a steady average travel time of approximately 80 to 81 seconds even as the traffic demand scales drastically from 250 up to 2,500 veh/h. This flat trend is consistently replicated in Manhattan, Rome, and Bologna scenarios.
In contrast, all baseline models exhibit a clear and rapid degradation in performance as traffic volume grows. The curves for these benchmarks escalate quickly and are truncated in the figures. 
This truncation occurs because, at sufficiently high traffic volumes, the baseline models can enter gridlock conditions in which mutually blocking vehicles prevent further progress through one or more intersections. The maximum traffic demand at which this occurs is highly scenario-dependent. Figure~\ref{fig:congestion} illustrates a representative localized gridlock event. Such events arise because reactive car-following and gap-acceptance mechanisms can lead to queues that propagate across adjacent junctions, eventually creating mutually blocking traffic streams. 
VeloCity is inherently robust to this problem. By leveraging a sequential distributed planning architecture, each vehicle computes its spatio-temporal trajectory in advance, preemptively reserving clear paths through complex intersections.

\begin{figure*}[t]
    \centering

    \includegraphics[width=\textwidth]{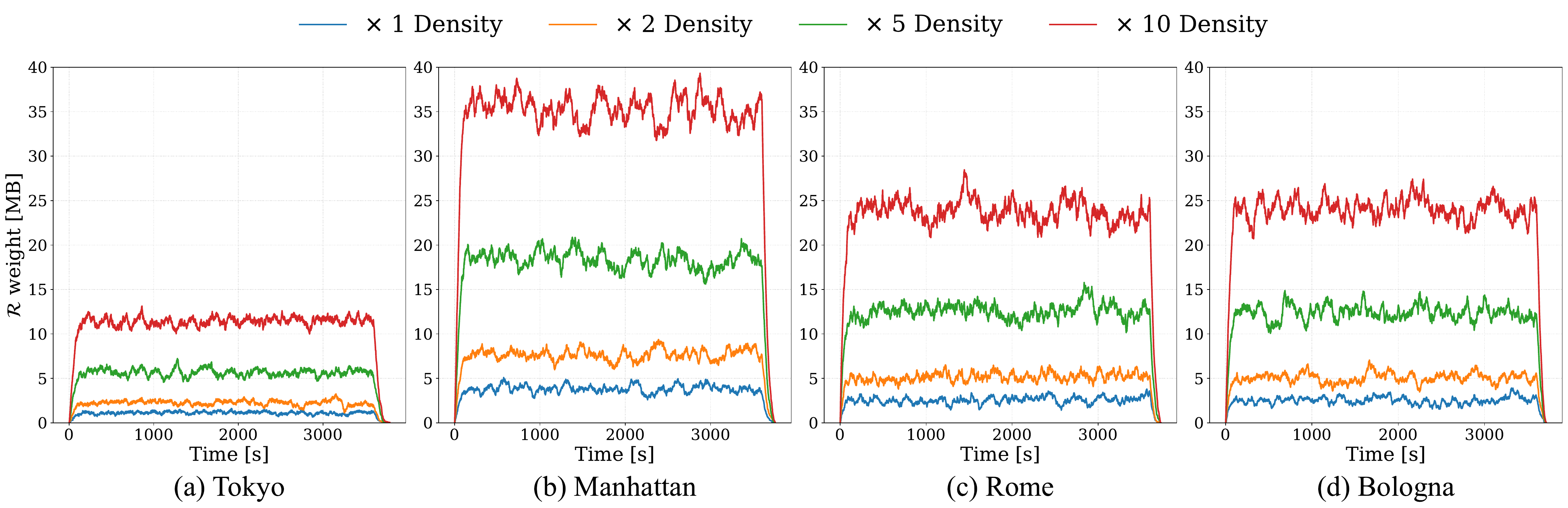}
    \caption{Reservation table $\mathcal{R}$ weight over time for the four evaluated urban scenarios, where the base density is (a) 250 veh/h for Tokyo, (b) 1,500 veh/h for Manhattan, (c) 6,000 veh/h for Rome, and (d) 2,500 veh/h for Bologna.}
    \label{fig:reservation}
\end{figure*}

\begin{figure*}[t]
    \centering

    \includegraphics[width=\textwidth]{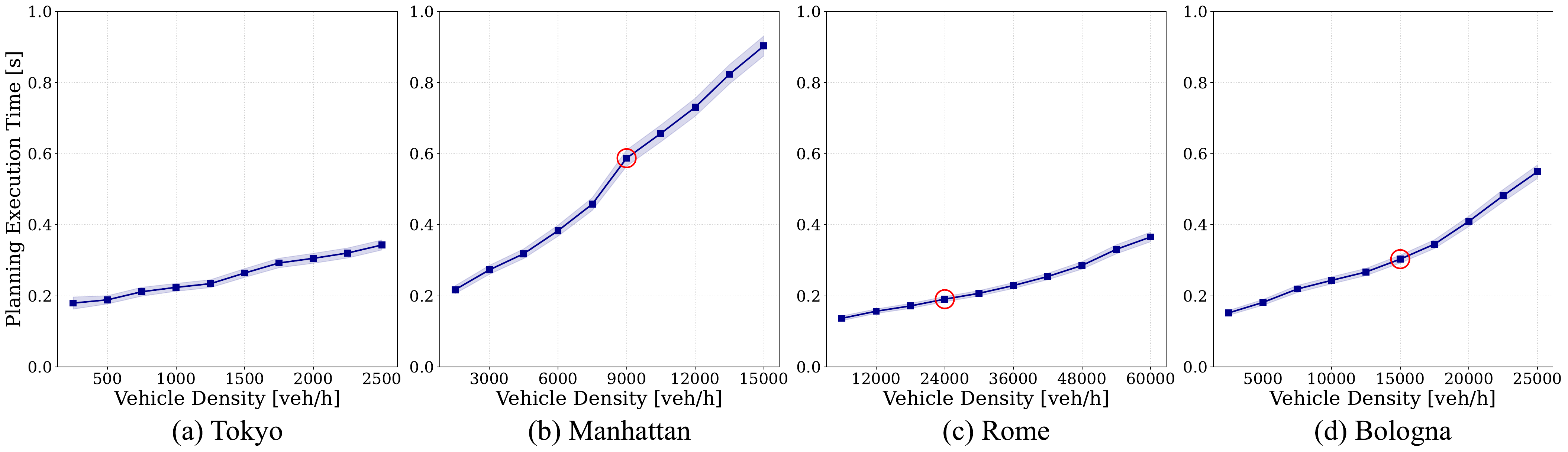}
    \caption{Vehicle planning time for the four evaluated urban scenarios.}
    \label{fig:planning_time}
\end{figure*}

Figure \ref{fig:box_plot} illustrates the actual distribution of travel times for each scenario at the maximum density safely supported by all baselines. The box plots reveal that VeloCity not only reduces the mean travel time, but it also reduces variance.
This confirms that VeloCity provides highly predictable and deterministic service guarantees.
Finally, in Fig.~\ref{fig:energy}, we evaluate the average per-vehicle energy consumption. Because VeloCity is designed to optimize travel times, vehicles traverse the network at higher average speeds. Therefore, the sustained higher cruising speeds can result in slightly higher overall energy consumption compared to the baselines.

\subsection{COORDINATION AND PLANNING ANALYSIS}




To ensure high computational efficiency and fast response times, the spatio-temporal reservation table $\mathcal{R}$ is implemented leveraging a hash map data structure, where each committed footprint $\mathcal{F}_i^k$ at time $t_k$ is mapped to a unique hashed key. 
Essentially, a hash map functions as a highly optimized dictionary that links unique identifiers (i.e., the hash keys) directly to stored data. In this context, the spatial coordinates and timestamp of a footprint are mathematically transformed into a single key, enabling the system to instantly check a specific location's availability in constant time ($\mathcal{O}(1)$). 
In this way, rather than maintaining a dense grid (which would scale poorly with the area size and time horizon), the coordinator stores only the actively reserved discrete space-time cells. Fig.~\ref{fig:reservation} illustrates the active memory footprint of the reservation table for VeloCity across the four evaluated scenarios, scaled at 1$\times$, 2$\times$, 5$\times$, and 10$\times$ of their respective baseline densities. The hash map representation proves to be lightweight: for the vast majority of cases, the memory utilization remains in the order of a few megabytes. The maximum memory footprint is reached only under the most extreme tested conditions, specifically during the maximum density peaks in the Manhattan scenario.

Beyond memory efficiency, we evaluate the system's temporal scalability by analyzing the total planning time required for a vehicle to be fully scheduled. As depicted in Fig.~\ref{fig:planning_time}, this total planning time encompasses three sequential steps: 1) the coordinator transmits the current reservation table $\mathcal{R}$ to the CAV entering the control area; 2) the CAV executes the local optimization outlined in Algorithm~\ref{alg:velocity_main}; and 3) the CAV transmits its computed reservation request $\mathcal{Q}_i$ back to the coordinator. For the communication overhead, we assume the deployment of standard 5G \ac{V2I} networks, which impose a maximum nominal latency of 10 ms for both the uplink and downlink transmissions \cite{moto2019field}. The local computation (Step 2) is evaluated over a standard single-thread Intel Core i7 processor.
Under these hardware and network assumptions, VeloCity successfully resolves the entire planning loop for each vehicle in just a few hundreds of milliseconds for most cases. For instance, in the Tokyo scenario, the mean planning time smoothly scales from roughly 180 ms at 250 veh/h to just 343 ms at 2,500 veh/h.
However, our stress-test experiments deliberately subject the framework to extremely high vehicular demands to identify its theoretical operational limits. Specifically, we push the Manhattan scenario up to 15,000 veh/h, Rome to 60,000 veh/h, and Bologna to 25,000 veh/h. To prevent a backlog of unscheduled vehicles, the per-vehicle planning time $t_{\text{plan}}$ must not exceed the average time between vehicle arrivals. Given a traffic density $D$ (in vehicles per hour), this constraint mandates that $t_{\text{plan}} \le (3.6 \times 10^6) / D$ (in ms). For example, a density of 1,000 veh/h yields a maximum planning window of 3,600 ms per vehicle.
While VeloCity comfortably satisfies this threshold for most traffic densities, executing Algorithm~\ref{alg:velocity_main} on a single thread eventually hits a computational bottleneck under extreme congestion. Notably, the adaption of dedicated hardware would significantly reduce planning times. Figure \ref{fig:planning_time} highlights with red circles the specific scenario-density pairs where the computational time exceeds the inter-arrival threshold, indicating the onset of queue accumulation. 
%
It is important to emphasize that this limitation stems purely from the baseline hardware assumptions and the current single-thread search implementation. 
One way to overcome this computational bottleneck could be extending Algorithm~\ref{alg:velocity_main} to a multi-threaded parallel search architecture, 
an enhancement previously proven highly effective for similar domains \cite{ahmadi2025parallelizing}. We leave the integration of parallelized local planning for future work.

\section{CONCLUSION}
\label{sec:conclusion}



This paper presented \emph{VeloCity}, a decentralized multi-agent spatio-temporal mobility profile planning framework designed to manage high-density automated traffic across arbitrary urban road topologies. VeloCity effectively bypasses the scalability bottlenecks of centralized traffic management systems by shifting from centralized mobility profile optimization to decentralized optimization directly solved at the individual \acp{CAV}.
Extensive simulations conducted in \ac{SUMO} across four large-scale real-world urban networks demonstrated the framework's superior efficiency. Compared to established baselines, VeloCity guarantees physically executable and collision-free mobility profiles while drastically reducing travel times, minimizing delay variance, and preventing cascading congestion even under high vehicular densities.

Despite these highly promising results, several avenues for future research remain. First, to further reduce the local planning time and eliminate computational queuing under ultra-high traffic demands, the underlying single-thread search algorithm can be extended into a multi-threaded parallel search architecture. Second, the reservation framework can be expanded to dynamically accommodate users with priorities (e.g., emergency vehicles). This would involve allowing priority entities to overwrite existing spatio-temporal allocations, requiring the development of a robust and efficient re-scheduling mechanism for affected \acp{CAV}. Third, future iterations can integrate cooperative mechanisms to safely handle mixed traffic scenarios, allowing non-cooperative or human-driven vehicles to coexist with the managed fleet. Finally, while the current objective function is tailored to strictly minimize system-wide travel time, incorporating energy consumption minimization into the local optimization problem will be crucial to explicitly balance the trade-off between crossing speed and overall energy efficiency.

\begingroup
\small
\bibliographystyle{IEEEtran}
\bibliography{references}
\endgroup

\begin{IEEEbiography}[{\includegraphics[width=1in,height=1.25in,clip,keepaspectratio]{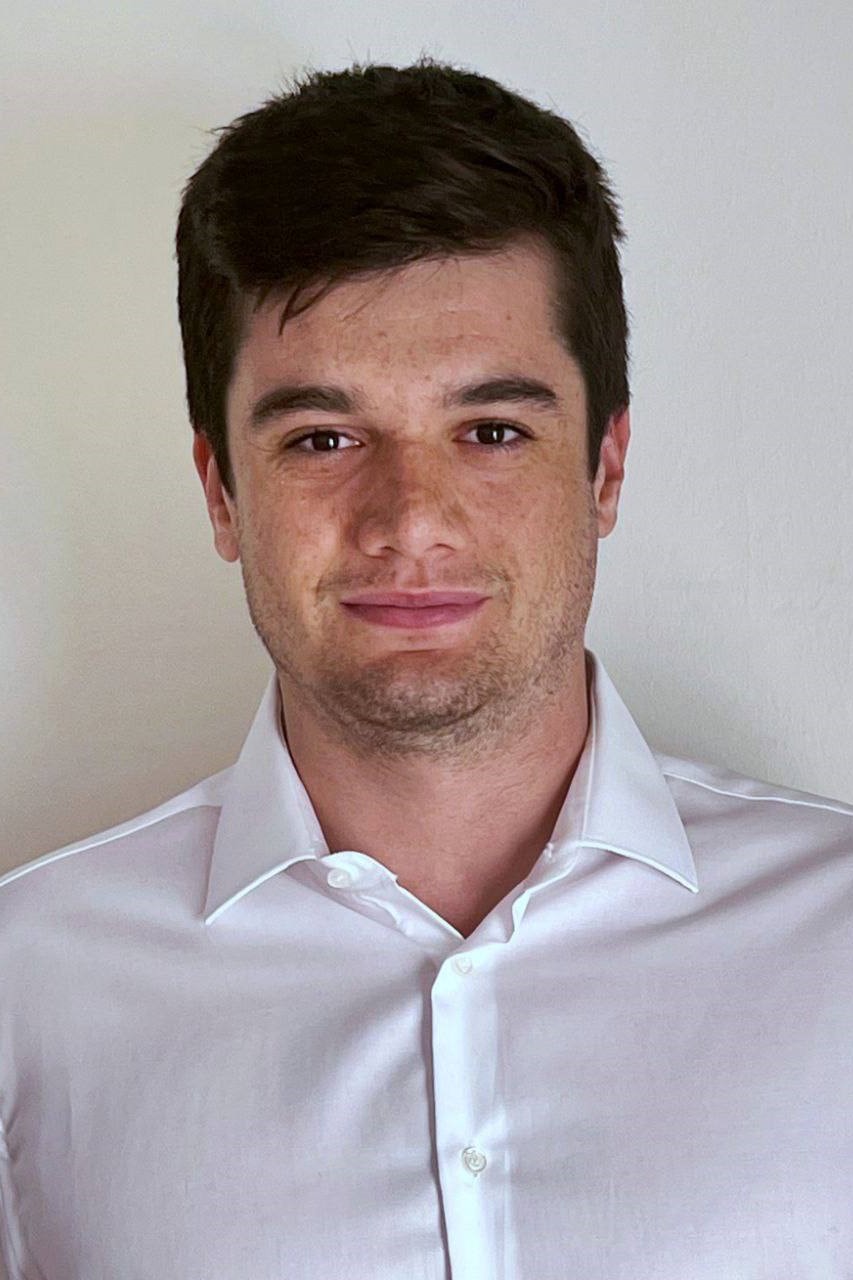}}]{Lorenzo Mario Amorosa }(Member, IEEE) received the B.S. degree in Computer Engineering and the M.S. degree in Artificial Intelligence from the University of Bologna, Italy, in 2019 and 2021, respectively. In 2025, he received the Ph.D. degree in Electronics, Telecommunications and Information Technologies Engineering from the University of Bologna. Currently he is with the Department of Electronic, Information and Electrical Engineering ``Guglielmo Marconi'' as a Postdoctoral Research Fellow at University of Bologna. He is Research Associate at the National Laboratory of Wireless Communications (WiLab) of CNIT (the National, Inter-University Consortium for Telecommunications). His main research interests include decentralized artificial intelligence, cooperative multi-agent systems, and deep learning for wireless communications.
\end{IEEEbiography}

\begin{IEEEbiography}[{\includegraphics[width=1in,height=1.25in,clip,keepaspectratio]{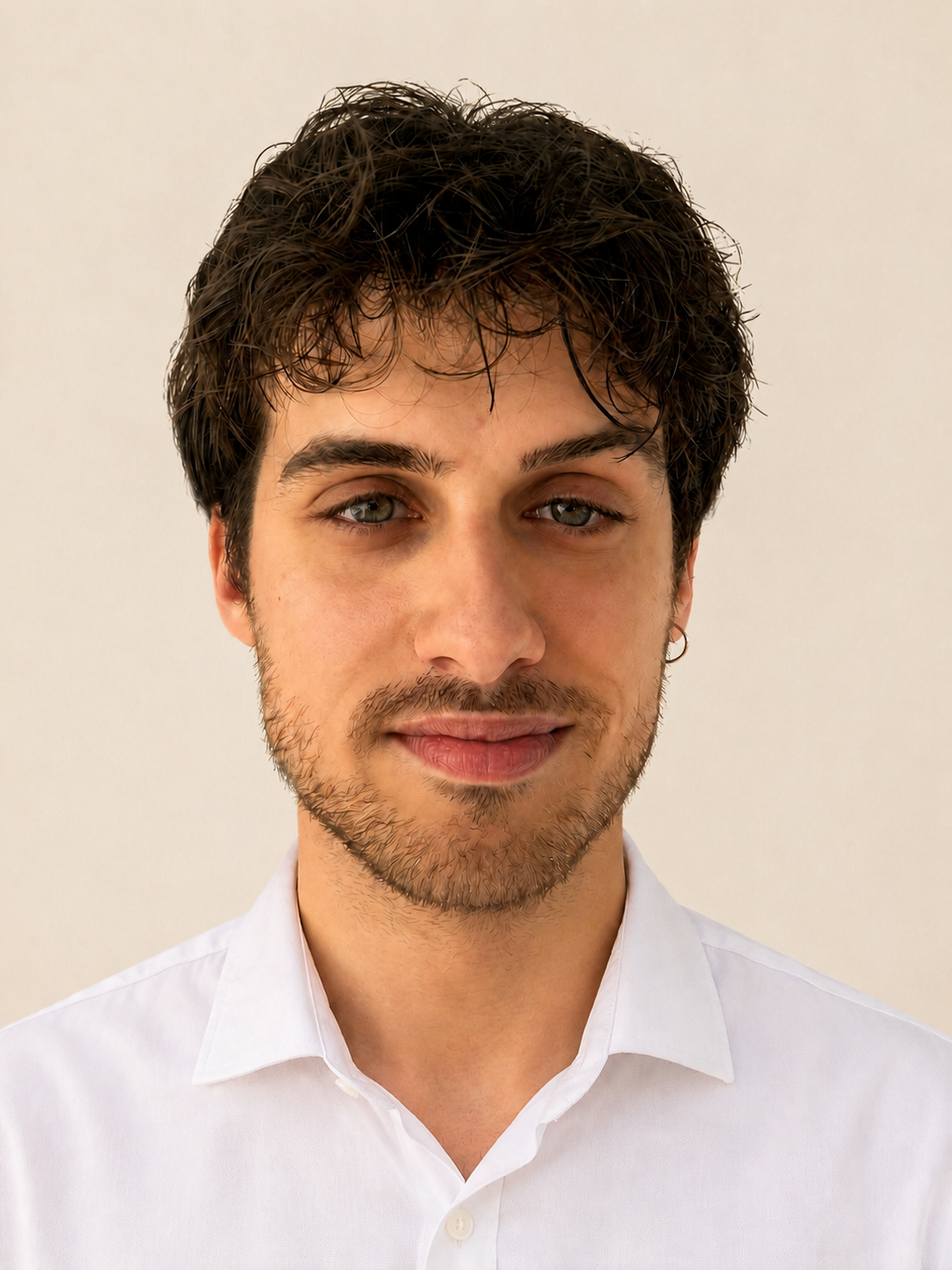}}]{Lorenzo Farina }(Graduate Student Member, IEEE) received the B.S. degree in Electronic and Telecommunications Engineering and the M.S. degree in Telecommunications Engineering from the University of Bologna, Italy, in 2021 and 2024, respectively. He is currently pursuing the Ph.D. degree in Automotive Engineering for Intelligent Mobility at the University of Bologna. He is Research Associate at the National Laboratory of Wireless Communications (WiLab) of CNIT (the National, Inter-University Consortium for Telecommunications). His research interests include networks of connected and autonomous vehicles, with a particular focus on algorithms and protocols for cooperative maneuvering.
\end{IEEEbiography}

\begin{IEEEbiography}[{\includegraphics[width=1in,height=1.25in,clip,keepaspectratio]{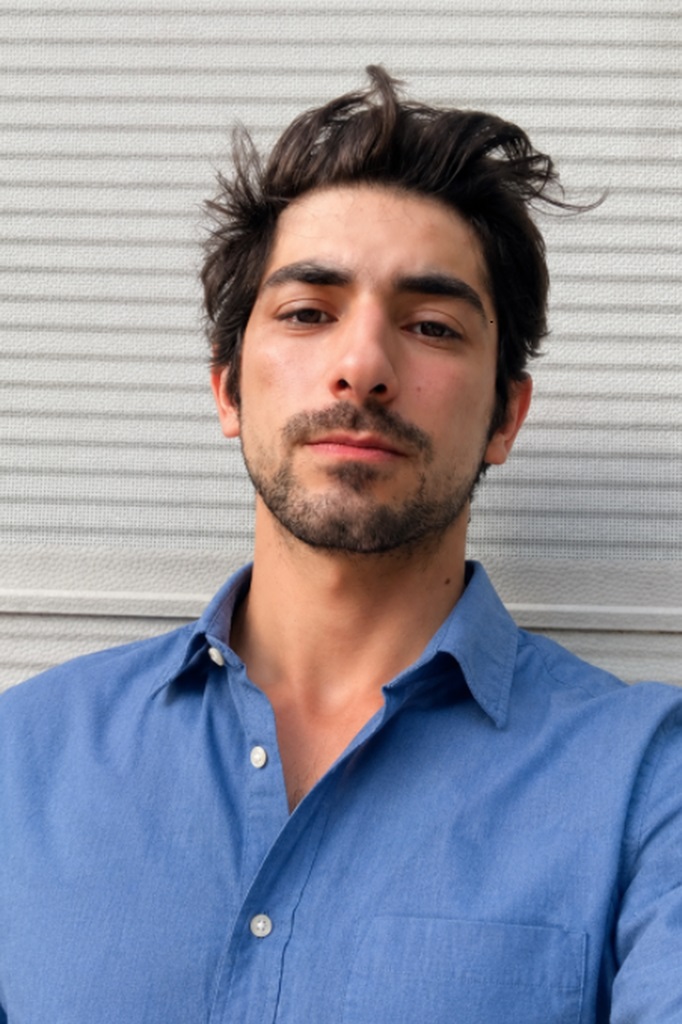}}]{Vittorio Todisco }(Member, IEEE) received the Ph.D. degree in Automotive Engineering for Intelligent Mobility from the University of Bologna, Italy, where he is currently a Research Fellow. His research focuses on wireless communications for connected and automated vehicles, including 5G/6G V2X, NR sidelink, distributed resource allocation, and interference management. He also closely follows ETSI standardization activities for intelligent transportation systems.
\end{IEEEbiography}

\begin{IEEEbiography}[\vspace{-0.2cm}{\includegraphics[width=1in,height=1.2in,clip,keepaspectratio]{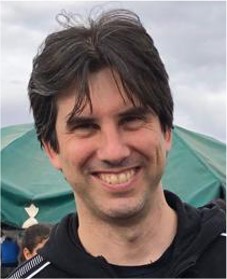}}]{Alessandro Bazzi}~(Senior Member, IEEE)  
is an Associate Professor at the University of Bologna and co-founder of the Wireless Communications Laboratory of CNIT, WiLab.  
His research interests focus on networks of connected and autonomous vehicles, including radio resource management of wireless networks and large-scale automotive radar interference. On these topics, he coordinates projects and contributes to ETSI standardization.
\end{IEEEbiography}

\end{document}